\documentclass[aps,prb,twocolumn,superscriptaddress,amsmath,amssymb,longbibliography]{revtex4-2}

\usepackage{bm}
\usepackage{amsmath}
\usepackage{amssymb}
\usepackage{graphicx}
\graphicspath{{img_1/}}
\usepackage{xcolor}
\input{epsf}
\usepackage[english]{babel}
\usepackage{gensymb}
\usepackage{ulem}
\usepackage[colorlinks,citecolor=cyan,linkcolor=cyan,urlcolor=cyan,bookmarks=false,hypertexnames=true]{hyperref}

\usepackage{booktabs}  

\begin{document}

\title{Driven-dissipative turbulence in exciton-polariton quantum fluids}

\author{R. Ferrini}
\affiliation{Center for Theoretical Physics of Complex Systems, Institute for Basic Science (IBS), Daejeon 34126, Republic of Korea}
\author{S.V.~Koniakhin}
\email{kon@ibs.re.kr}
\affiliation{Center for Theoretical Physics of Complex Systems, Institute for Basic Science (IBS), Daejeon 34126, Republic of Korea}
\affiliation{Basic Science Program, Korea University of Science and Technology (UST), Daejeon 34113, Republic of Korea}

\begin{abstract}
The present paper is devoted to comprehensive theoretical studies of exciton-polariton quantum fluids specificities in the optics of their utilization for quantum turbulence research. We show that a non-trivial implementation of time-varying potential for excitation of quantum fluid (injection of quantized vortices) via the stirring procedure can be efficiently substituted with resonant excitation-based phase-imprinting techniques. The most efficient phase pattern corresponds to imprinting of tiles with randomly oriented plane waves in each. The resulting turbulent flows, spatial vortex distributions, and clustering statistics resemble those for the case of a conventional spoon-stirring scheme. We quantify the limitations on the lifetime and density depletion for the development and sustainability of quantum turbulence. The yield is the necessity to prevent the density depletion for more than one order of magnitude. Finally, we demonstrate that turbulence is robust with respect to alternating gain and loss at a certain range of modulation parameters, which corresponds to laser operating above and below condensation threshold.


\end{abstract}
\maketitle

\section{Introduction}\label{sec:intro}
Quantum turbulence is a peculiar stochastic phenomenon characterized by the emergence of a spontaneous order from chaotic motion, due to the formation of self-similar spatial structures of vorticity topological defects in a quantum fluid. Contrary to the observations of 3D quantum turbulence in quantum systems, such as superfluid helium \cite{vinen1957mutual, skrbek2011quantum} and ultra-cold atomic Bose-Einstein condensates \cite{tsubota2013quantum, white2014vortices}, the existence of an inverse cascade for 2D quantum turbulence analogous to the one observed in classical fluids is still under the debate, due to both controversial scaling arguments and very stringent mathematical limits on numerical simulations \cite{tsubota2017numerical}. Indeed, while several works have numerically confirmed the dynamical clustering of quantized vortices, namely the Onsager vortex clustering, consistently with the inverse energy cascade by means of evaporating heating process~\cite{simula2014emergence,valani2018einstein, horng2009two,billam2015spectral}, other argue against its existence \cite{numasato2010direct}. Today, thanks to recent technological advancements in spectroscopy and semiconductor manufacturing, polariton quantum fluids, with their hybrid light-matter nature, represent an alternative platform for investigating 2D quantum turbulence \cite{berloff2010turbulence,panico2023onset}.


The exciton-polariton is a bosonic composite quasiparticle, formed in semiconductor microcavities by the exciton (electron-hole hydrogen-atom-like quasiparticle) strongly coupled to the cavity photon, and undergoes a strong radiative decay because of the finite lifetime of its photonic component. Compared to the long-living atomic systems, this essential feature, in addition to the very small effective mass of the polariton, distinguishes the observed superfluid behavior \cite{carusotto2004probing, amo2009superfluidity} and Bose-Einstein condensation \cite{kasprzak2006bose} in polaritonic systems. The latter exhibit a high critical temperature (very small effective mass) and a driven-dissipative nature, which experimentally can be overcome by means of external optical incoherent pumping \cite{roumpos2011single}. The dynamics of topological defects has been extensively investigated. The experimentally estimated healing length value $\xi$ (order of a micrometer) of quantized vortices ~\cite{lagoudakis2008quantized, dominici2015vortex, lagoudakis2011probing} allows, in principle, the opportunity to observe an inverse energy cascade over a larger range of magnitude of the wave vector (upper cascade bound $1/\xi$). Moreover, the quantized vortex creation method (in other words, the energy injection method) plays a crucial role in the observation of the energy cascade both theoretically and experimentally, and techniques such as spoon-like stirring \cite{madison2000vortex,reeves2012classical, reeves2022turbulent}, paddle, or defect grid drag \cite{gauthier2019giant,johnstone2019evolution} of fluid flowing around stationary defects \cite{reeves2013inverse} have been used in the past as a standard vortex injection technique for atomic systems. 

The necessity to implement a time-varying potential challenges the implementation of conventional spoon-stirring quantized vortex injection scheme for exciton polariton systems. Currently, among the setups based on the time-varying potential for polariton quantum fluid excitation, one should highlight the creation of one or several vortices using the potentials with rotation-induced chirality~\cite{del2023optically,gnusov2023quantum,gnusov2024vortex}. The experimental setups required combination of several lasers with finely controlled frequency difference to make the time-dependent interference-governed gain and potential landscape. The resulting potentials were smooth (and thus less effective in stirring) as a result of the smooth phase variation of the elemental beams. Finally, the long-living (hundreds of picoseconds) exciton reservoir~\cite{de2014relaxation,park2025exciton} will produce the unnecessary trail. However, polaritons are suitable for direct wave function imprinting via the quasi-resonant excitation approach with flexible spatial light modulator-based (SLM) phase engineering. This approach was successfully used to excite complex spatial distributions of quantized vortices~\cite{boulier2015vortex,panico2021dynamics}. It is noteworthy that the atomic condensates can also be subjected to the phase imprinting techniques realized by the pulsating potential for e.g. the soliton creation~\cite{burger1999dark}. It is also necessary to find reliable approaches for that type of excitation that are able to reproduce the specificity of conventional spoon stirring in the framework of vortex clustering and spectral energy distribution. In this study, we propose various wave function phase patterns and show that those of tiles of plane waves with randomly directed wave vectors effectively mimic the spoon-stirring method of vortex injection. 

Next, the important specificity of polaritons is finite lifetime. Previously, the effects of the finite lifetime of polaritons were investigated in Ref.~\cite{comaron2024dynamics} with the main emphasis on tracking the time dynamics of vortex spacing, dipole moments and vorticity correlation function. It remains important to study incompressible kinetic energy spectra during the decay process, and this study covers this gap.

Finally, we investigate the effects of gain/loss fluctuations on the reliable existence of quantum turbulence. First, we impose experimental limitations on the required stability of laser power to maintain the turbulence by means of an incoherent excitation scheme. At the same time, we show that alternating periods of positive and negative gain under certain conditions do not affect the development of the turbulence. Thus, an experimental scheme of the laser periodically operating above and below the condensation threshold can be useful for quantum turbulence experiments: the positive gain phase overcomes the losses, and the negative gain phase provides the unperturbed wave function evolution.

The paper is organized as follows. In section~\ref{sec2_setup}, we describe in detail the scheme of the provided numerical simulation of dissipative quantum fluid dynamics within the framework of Gross-Pitaevskii equation. In subsection~\ref{subsecIIIa_comparison_strategies}, the equivalence between imprinting techniques and spoon stirring for quantum turbulence studies is extensively discussed within the framework of spatial distributions, vortex statistics and incompressible kinetic energy (IKE) spectra analysis. Subsection~\ref{subsecIIIb_losses} focuses on
the effects of losses on sustainability of quantum turbulence arising from the finite lifetime specificity of polaritons, 
studying at which extent quantum fluids density depletion effects on IKE spectra and vortex statistics for quantum fluid with losses. Then, in subsection~\ref{subsecIIIc_gain/loss_fluct}, we investigate the conditions on the existence of reliable quantum turbulence signs in presence of gain/loss fluctuations, modeling the experimental case of a laser operating above and below condensation threshold. The final conclusions and observation are collected in section~\ref{secIV_conslusion}.

\section{Numerical setup}
\label{sec2_setup}
In the quantum fluids research field, the widely used theoretical model is the mean-field Gross-Pitaevskii equation-based approach, which efficiently describes a Bose-Einstein condensate in equilibrium. Regarding exciton-polaritons, quasiparticles of finite lifetime, such a model represents an idealized conservative description and requires modification by adding the specific terms responsible for the gain and loss:
\begin{equation}
    \label{eq: modified GPE}
    \begin{split}
    i\hbar\frac{\partial}{\partial t} \psi(\mathbf{r}, t) = &\left(i\beta(t)-1\right)\frac{\hbar^2\nabla^2}{2m}\psi(\mathbf{r},t)+g|\psi(\mathbf{r},t)|^2\psi(\mathbf{r},t)\\& + V(\mathbf{r},t)\psi(\mathbf{r},t)-i\frac{\hbar}{2}\gamma(t) \psi(\mathbf{r},t).
    \end{split}
\end{equation}

The values for a polariton system are taken as follows: $m=5\times10^{-5} \ m_0$ (bare electron mass $m_0= 9.1\times10^{-28}$~g), $g = 5\times10^{-3}~\mathrm{meV}\mu \mathrm{m}^2$. Eq.(\ref{eq: modified GPE}) is solved by means of third order Adam-Bashforth method for the time integration, with time step $\Delta t=0.004$~ps, total duration $T_{tot}=4.56$~ns and a square mesh grid of size $L=512 \ \mu$m. For the calculation of the Laplace operator, we used the Fourier transform with massive parallelization provided by GPU. The additional $\beta$-dependent term comes from the hybrid Boltzmann Gross-Pitaevskii theoretical model~\cite{Solnyshkov2014}, where the $\beta$ parameter governs the dissipation related to high wave vectors. The value of $\beta$ parameter is predicted to be controlled by the temperature from nearly zero value to that comparable to 0.1. 

 In this study, as one of the principal goals, we present the turbulence generation methods based on direct phase imprinting (governed by the initial wave function $\psi_0=\psi(\mathbf{r}, t=0)$) and compare them with the spoon stirring as a reference. For all methods, the simulation of the total duration $T_{\mathrm{tot}}=4.56$~ns was divided into the preparatory part and the analysis part. The confining circular hard wall potential of 480~$\mu$m in diameter was normally used.  During the first (preparatory) part of the simulation (from 0 to $T_{\mathrm{tot}}/2 = 2.28$~ps), polariton lifetime was set sufficiently higher than simulation duration and decay parameter $\beta$ was set to $0.001$ to suppress the short wave length density waves that complicate the vortex detection. During the second (analysis) part of the simulation from $T_{\mathrm{tot}}/2$ to $T_{\mathrm{tot}}$, the data for the analysis was collected. The parameter $\beta$ was switched to zero and lifetime $\gamma^{-1}$ was set depending on the required decay rate. We use the values of polariton lifetime listed in Table~\ref{tab_1}, which yield the final density 30\%, 10\%, 3\% and 1\% of the initial one (for the analysis phase). These values exceed sufficiently the achievable in actual samples, however they allow establishing fundamental limitations for the polariton platform utilization for quantum turbulence studies. The duration of the spoon-stirring numerical experiment was similar, with stirring phase during the first part and analysis during the second part. More simulation details, including the videos, are given in the Appendix.

\begin{figure}
    \begin{center}
        \includegraphics[width=\linewidth]{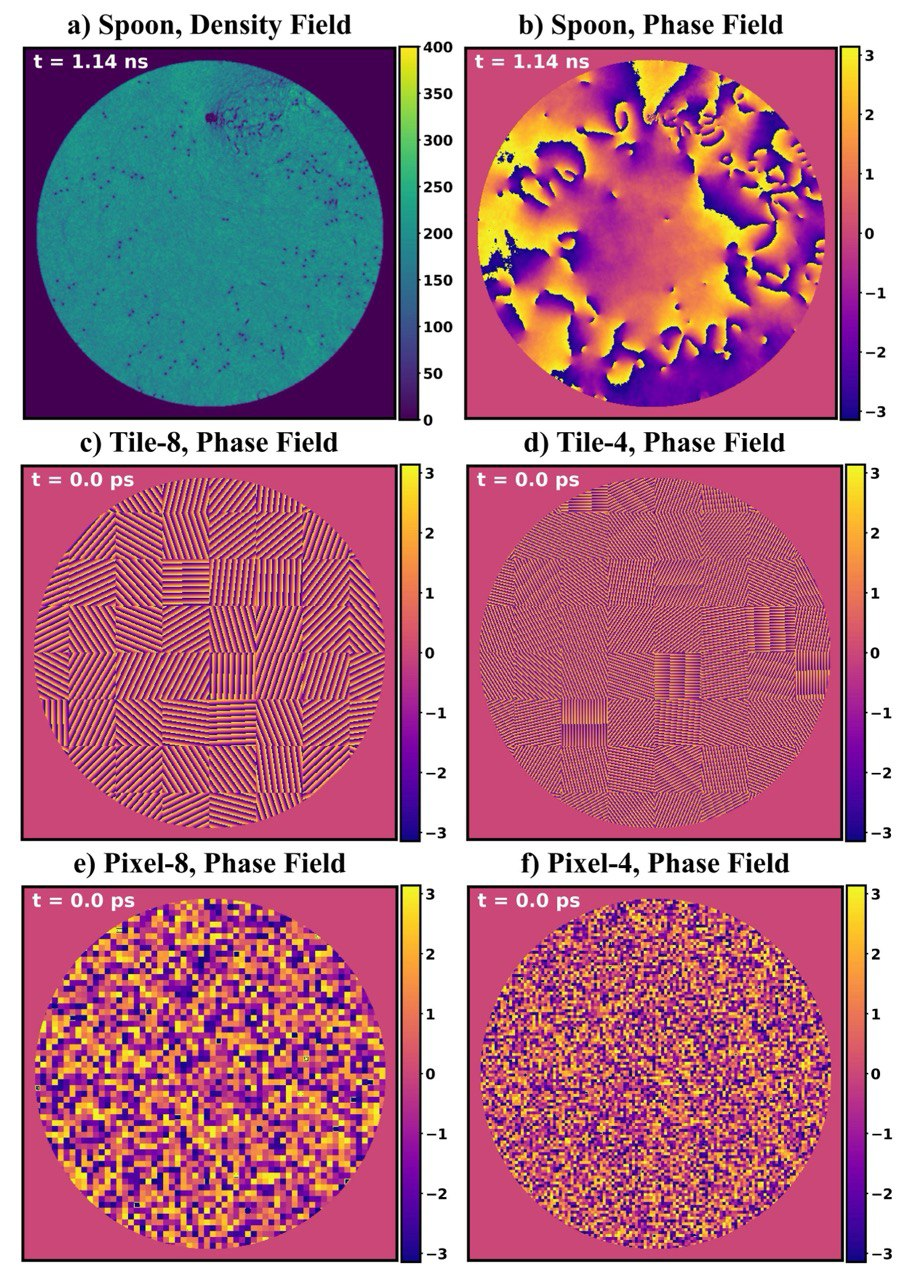}
    \end{center}
    
    \caption{\label{fig_1_stirr} Snapshots of condensate wave function to illustrate the various employed quantum fluid excitation schemes. For the rotating-spoon method, density and phase fields are respectively shown in panels \textbf{a)} and \textbf{b)} (one can see the spoon potential stirring the initially homogeneous system). While, for the imprinting-strategies the characteristic field is the wave function phase at initial time moment $t=0$~ps. Tile-imprinting strategies (with 8x8 grid of $32~\mu$m tiles) are illustrated in \textbf{c)} for Tile-8 scheme (plane wave wavelength $\lambda = 8~\mu$m) and in \textbf{d)} for Tile-4 ($\lambda = 4~\mu$m). Finally, snapshots \textbf{e)} and \textbf{f)} show the initial imprinted phase pattern for pixel-imprinting technique, with pixel sizes 8~$\mu$m (Pixel-8) and 4~$\mu$m (Pixel-4) respectively }
     
\end{figure}

The simulation of spoon stirring was carried out by a 10~$\mu$m potential propagating with a speed of 1.3~$\mu$m/ps during the preparatory phase. As for the imprinting methods, the two different families of phase pattern were considered. In the first case, we divide the wave function into the 8x8 grid of the square tiles of size 64~$\mu$m and fill each of them with a randomly oriented plane wave with wave length 12~$\mu$m (Tile-12), 8~$\mu$m (Tile-8) or 4~$\mu$m (Tile-4). In the second case, we divide wave function into pixels of the size 8~$\mu$m (Pixel-8) or 4~$\mu$m (Pixel-4) of the random phase each. In both cases, we keep the density of the wave function at the level of approximately $120~\mu$m$^{-2}$ which gives healing length $\langle\xi(t)\rangle=\hbar\sqrt{2m\langle|\psi(t)|^2\rangle}\approx 1.2 ~\mu$m. Experimentally, the phase imprinting can be straightforwardly implemented based on the quasi-resonant pumping and spatial light modulator usage as it was done previously for vortex chains~\cite{boulier2015vortex} and lattices~\cite{panico2021dynamics}. By the principles, the tile imprinting schemes resemble the orbital angular momentum injection by four arranged in square lasers described in~\cite{boulier2016injection} For spoon-stirring and tile-imprinting turbulence generation schemes, the density in the beginning of analysis phase was around 120~$\mu$m$^{-2}$, while for pixel schemes one obtains around 160~$\mu$m$^{-2}$. Fig.~\ref{fig_1_stirr} illustrates the basic principles of the investigated vortex injection and turbulent motion excitation methods. Practically, for the imprinting schemes, the preparatory phase (where density waves relaxation and turbulent regime stabilization takes place) could be significantly lower, i.e. around 200~ps. See the simulation videos in the Appendix. This is yet another advantage of the imprinting techniques with respect to the ``spoon stirring''.

\begin{center}
\begin{table}[]
\centering
\vspace{3em} 
\caption{\label{tab_1}Quantum fluids densities $\langle|\psi|^2\rangle$ in the end of analysis phase with respect to initial one used for simulation and corresponding values of lifetime.}
\begin{tabular}{|c|c|c|c|c|c|}
\hline
$\dfrac{\langle|\psi|^2\rangle(t=T_{\mathrm{tot}})}{\langle|\psi|^2\rangle(t=T_{\mathrm{tot}}/2)}$ 
& 0.3 & 0.1 & 0.03 & $\approx 1$ \\
\hline
$\tau$ (ns) & 1.89 & 0.99 & 0.65 & 300 \\

\hline
\end{tabular}

\end{table}
\end{center}

For the analysis, we perform vortex detection and implement clustering algorithm described in section VI of Ref.~\cite{valani2018einstein}. It allows calculating the number of free and clustered vortices and of vortex dipoles in each time moment. Knowing the positions of vortices (and chemical potential $\mu = gn$) allows \textit{semi-analytical} calculation of IKE spectra both for full vortex distributions and for clustered vortices only, giving priority to hierarchal structures specific to the turbulence with respect to single vortices and dipoles~\cite{Koniakhin2020}. The function $E^{(i)}(k)$ representing the IKE spectrum we are interested in can be calculated as
\begin{equation}
    \label{eq: analyt_IKE}
    E^{(i)}(k)= N_{vort}\Omega\xi^3F(k\xi)G(k),
\end{equation}
where
\begin{equation}
    F(k\xi) = \frac{k\xi}{4 \Lambda}\left(I_1\left(\frac{k\xi}{2 \Lambda}\right)  K_0\left(\frac{k\xi}{2 \Lambda}\right) -I_0\left(\frac{k\xi}{2 \Lambda}\right)  K_1\left(\frac{k\xi}{2 \Lambda}\right)   \right) ,
\end{equation}
and vortex positions and signs define
\begin{equation}
    G(k)=1+\frac{2}{N_{vort}}\sum_{i=1}^{N_{vort}-1}\sum_{j=i+1}^{N_{vort}} \kappa_i\kappa_jJ_0(k|\mathbf{r}_i-\mathbf{r}_j|){}
\end{equation}
with ensthropy quantum $\Omega=2\pi\hbar^2n/(m\xi^2)$, parameter $\Lambda=0.8249$.

We also perform calculating the IKE spectra fully \textit{numerically}. First, the density weighted velocity field $u(\mathbf{r})= -i\hbar/(2m {|\psi(\mathbf{r})|)} \cdot(\psi^*\left(\mathbf{r})\nabla\psi(\mathbf{r})-\psi(\mathbf{r})\nabla\psi^*(\mathbf{r})\right)$ is calculated and plugged into equation for finding the incompressible $(i)$ part:
\begin{equation}
    u_{\alpha}^{(i)}(\mathbf{k})=\sum_{\beta=x,y}(\delta_{\alpha,\beta}-k_{\alpha}k_{\beta}/k^2)u_{\beta}(\mathbf{k}),\ \text{for}\ \alpha=x,\ y.
\end{equation}
The latter, is then used to calculate the IKE spectrum via
\begin{equation}
\label{eq: IKE general formula}
        E^{(i)}(k) = \frac{m}{2}\int \ d\mathbf{k} \left(|u_x^{(i)}(\mathbf{k})|^2+|u_y^{(i)}(\mathbf{k})|^2\right).\\
\end{equation}

Furthermore, by means of Eq. \eqref{eq: modified GPE} with time-dependent loss rate $\gamma(t)$, we present a model with sign-alternating polariton gain and loss, which approximates the behavior of the driven-dissipative Gross-Pitaevskii equation (ddGPE) for polariton condensate incoherently pumped by the unstable or intentionally modulated laser operating above and below the threshold. For this set of simulations, we do not employ the confining potential.

We averaged the vortex statistics and spectra over 16 runs for each set of parameters. For the tile-imprinting we changed randomly the wave vector directions within the tiles, for pixel-imprinting we change the pixel phases and for the spoon-stirring we slightly change the rotation orbit diameter.

\section{Results and discussions}

\subsection{Comparison of imprinting techniques and spoon-stirring vortex injection}
\label{subsecIIIa_comparison_strategies}

The main goal of present subsection is to clarify if imprinting techniques are equivalent to spoon stirring for quantum turbulence studies. Therefore, we consider the data at the very initial time moments of the analysis phase, i.e. from $T_{\mathrm{tot}}/2 = 2.28$~ps to 2.38 ps and average over 10 time moments. We use the resulting wave functions and compare vortex and cluster statistics and IKE spectra for reference spoon stirring procedure of turbulent motion excitation and for that relying on imprinting techniques.

\begin{figure}
    \begin{center}
        \includegraphics[width=\linewidth]{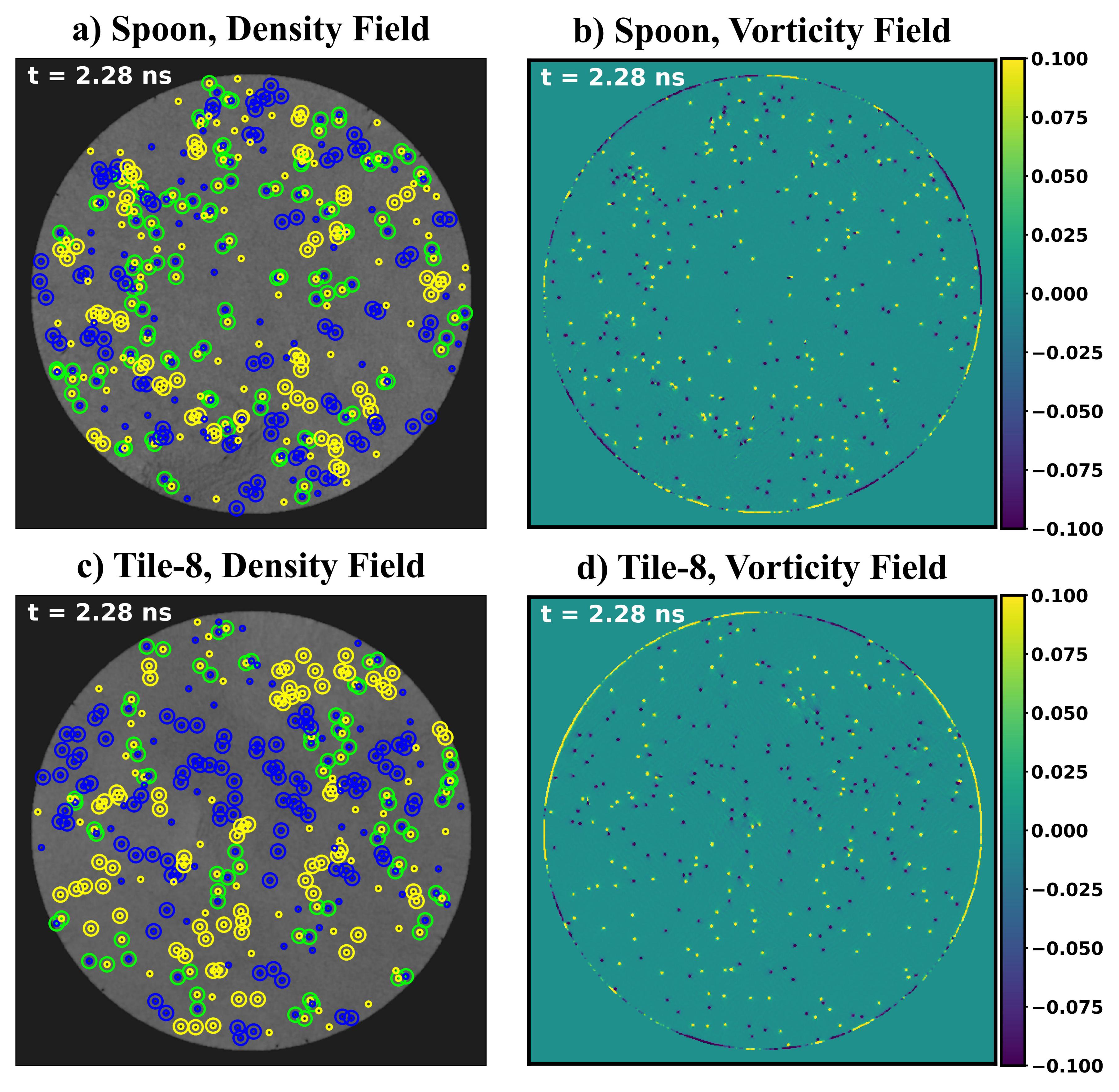}
    \end{center}
    
    \caption{\label{fig_2_detection} Results of vortex detection followed by clustering algorithm for spoon-stirring \textbf{(a,b)} and Tile-8 \textbf{(c,d)} excitation. The vortices of opposite signs are denoted with yellow and blue inner circles in density field images (panels \textbf{a} and \textbf{c}). Vortices belonging to the clusters carry outer circles of the corresponding colors. Green outer circles show the dipoles. Vorticity field is shown in panels \textbf{b)} and \textbf{d)}.
    }
\end{figure}

Fig. \ref{fig_2_detection} shows the results of vortex detection and the application of the clustering algorithm for the cases of the Tile-8 phase-imprinting technique and for the spoon-stirring procedure at the beginning of the analysis phase (for conservative simulation). The more detailed simulation snapshot series and links to the videos are given in Appendix. For the spoon-stirring, the chosen simulation time was enough to generate significant amount of quantized vortices and their clusters. For the tile-imprinting, during the first several dozens of picoseconds, stabilization of vortices seeded by the initial phase singularities occurred. Further, the long-wavelength density waves extinction and quantum fluid homogenization took place. Visually, the resulting vortex distributions and their organization into clusters nearly coincide for Tile-8 and spoon-stirring. However, a detailed quantitative analysis of the excitation strategies is necessary from the point of view of cluster statistics and IKE spectra.

\begin{figure}
    \begin{center}
        \includegraphics[width=\linewidth]{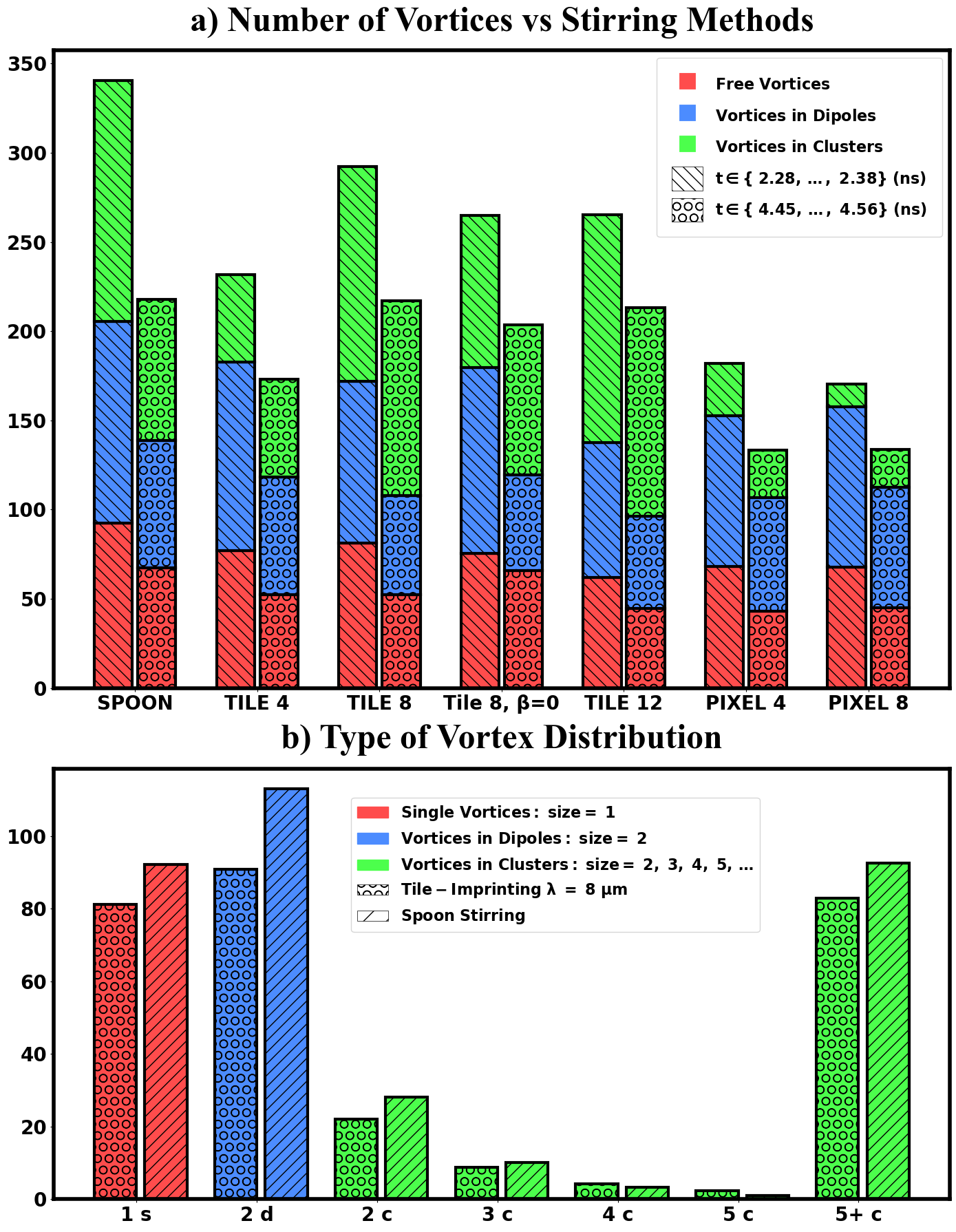}
    \end{center}
    
    \caption{\label{fig_3_hist} Histograms for vortex statistics results obtained for all excitation strategies. Panel \textbf{a)} is shows single vortices (red bar parts), vortices in dipoles (blue) and clustered vortices (green) in the beginning of the analysis phase (left of paired bars) and in the end of the analysis phase (right of paired bars). Particle lifetime is set $\tau=1.89\ ns$. Panel \textbf{b)} focuses on Tile-8 phase imprinting strategy and spoon-stirring strategy giving details on the number of vortices belonging to clusters of various sizes. The data in panel \textbf{b)} corresponds to the beginning of the analysis phase and thus to the left counterparts of the paired columns in panel \textbf{a)}.}
\end{figure}

To do so, we plot histograms to highlight vortex clustering for all excitation strategies employed. The left counterparts of the paired columns in panel a) in Fig.~\ref{fig_3_hist} show the number of single vortices, vortices in dipoles and vortices in clusters at the beginning of the analysis stage for all excitation techniques employed. One can see that Tile-8 and Tile-12 strategies resemble the spoon-stirring one much better than other strategies. Fractions of vortices in single state, dipoles and clusters are sufficiently similar. Due to this coincidence, Tile-8 with tile size of 32~$\mu$m (see Fig. S3 for alternative grids) can be chosen as a reference phase-imprinting method for more detailed comparison with the spoon stirring approach. Setting decay parameter $\beta$ to zero during the preparatory part had rather quantitative than qualitative effect as it is shown in Appendix. For Tile-4 and Pixel strategies, one observes substantially lower fraction of vortices in clusters. The pixel schemes do not provide the global rotational motion and Tile-4 scheme injects too high kinetic energy with respect to potential one. Panel b) in Fig.~\ref{fig_3_hist} gives more details on clustering for spoon stirring and phase imprinting Tile-8 strategy. Tile-8 gives a slightly higher number of vortices in large clusters whereas the spoon-stirring strategy produces more single vortices, dipoles and small clusters. Typical intervortex distance $l$ can be estimated as $L/\sqrt{N}$, which gives $l\approx 10~\mu$m for Tile-8 and spoon-stirring strategies. The dominant fraction of clustered vortices is concentrated in large clusters.

\begin{figure}
    \begin{center}
        \includegraphics[width=\linewidth]{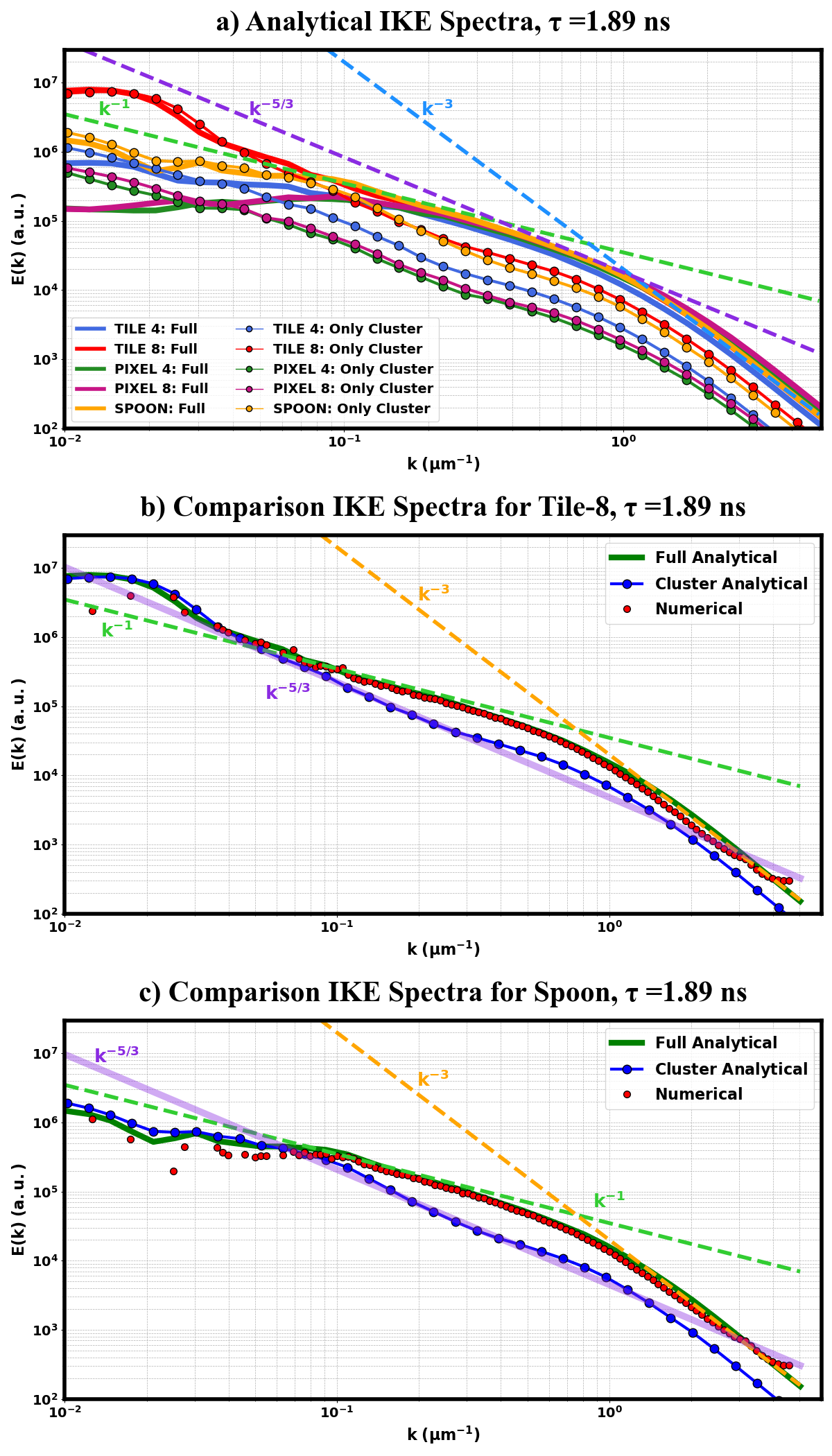}
    \end{center}
    
    \caption{\label{fig_4_IKE3} Panel \textbf{a)}. Results of the Incompressible Kinetic Energy (IKE) spectra semi-analytical calculation (Eq.~\eqref{eq: analyt_IKE}) for all excitation strategies at the beginning of the analysis phase (approximately 2.3~ns). Solid curves are for full vortex distributions and curve with dots is for clustered vortices only. Guides for eye are given for characteristic exponents: -3 (vortex core), -1 (dipole), -5/3 (clustered vortices). Panels \textbf{b)} presents the comparison of full IKE spectra obtained semi-analytically and numerically and IKE spectra obtained semi-analytically for clustered vortices for Tile-8 excitation. Panel \textbf{c)} is the same for the spoon-stirring approach.
    }
\end{figure}

IKE spectra calculated via positions of all vortices and via positions of clustered vortices only (employing Eq.~\eqref{eq: analyt_IKE}) are depicted in panel a) of Fig.~\ref{fig_4_IKE3} for all excitation strategies. One can see very similar full IKE spectra regardless of the strategy employed; the discrepancies are within the values imposed by the difference in the total number of vortices, see Fig.~\ref{fig_3_hist}. The full IKE spectra are composed of two parts different in characteristic power law. Exponent -3 corresponds to vortex cores at high wave vectors (greater than $\xi^{-1}$) and it is presented in IKE spectra for all excitation methods and analysis frameworks. The exponent -1 dominates at lower wave vectors, being a fingerprint of an uncorrelated gas of vortices with dominating dipoles and single vortices. At the same time, the spectra for clustered vortices demonstrate a visible region of -5/3 power law, as was shown earlier for different excitation strategies~\cite{Koniakhin2020}. This region spans from $k_p$ to $l^{-1}$, where the high-scale boundary is defined by a tile size as $k_p \approx 2/l_T$. The system has a certain memory effect of the tile size $l_T$ visible by the energy surplus stored at the wave vector $k_p$, see Appendix, Fig. S4, S5. From $l^{-1}$ to $\xi^{-1}$ the slope -1 takes place. It should be noted that the Tile-8 phase-imprinting excitation strategy demonstrates full and clustered vortex-based IKE spectra very similar to those of the spoon-stirring approach. Among the phase-imprinting techniques, Tile-8 yields the highest clusterization degree comparable to the spoon stirring case. For pixel-type imprinting contribution from clustered vortices is lower, which agrees with the histograms. Panels b) and c) in Fig.~\ref{fig_4_IKE3} show full IKE spectra obtained semi-analytically via Eq.~\eqref{eq: analyt_IKE} (for full IKE spectra and IKE spectra of clustered vortices) and numerically via Eq.~\eqref{eq: IKE general formula} for Tile-8 and spoon stirring cases, respectively. One sees that at low wave vectors (large size scales) full and clustered vortex IKE spectra coincide. This is due to the vanishing global rotational motion from the uncorrelated part of the ``vortex gas''. For the wave vectors above $l^{-1}$, one sees that full IKE spectra have higher magnitude than the clustered vortex IKE spectra proportionally to the fraction of clustered vortices. The compressible energy contribution is lower than that of the incompressible, see Fig. S5.


The main conclusion of present section is that the phase-imprinting excitation strategy Tile-8 gives vortex statistics and IKE spectra very close to those of the conventional spoon stirring excitation strategy. Thus, it could be used as a phase pattern sent to SLM for quasi-resonant polariton excitation in experiments.  The proposed Tile approaches provide sufficient degree of stochasticity due to random orientations of the flow directions within different tiles. Therefore, one will be able doing both reproducible (for at least initial evolution stages) runs for a phase given pattern for emission signal accumulation and perform averaging over numerous random initial phase patterns.


\subsection{Effects of losses on sustainability of quantum turbulence}
\label{subsecIIIb_losses}

\begin{figure}
    \begin{center}
        \includegraphics[width=\linewidth]{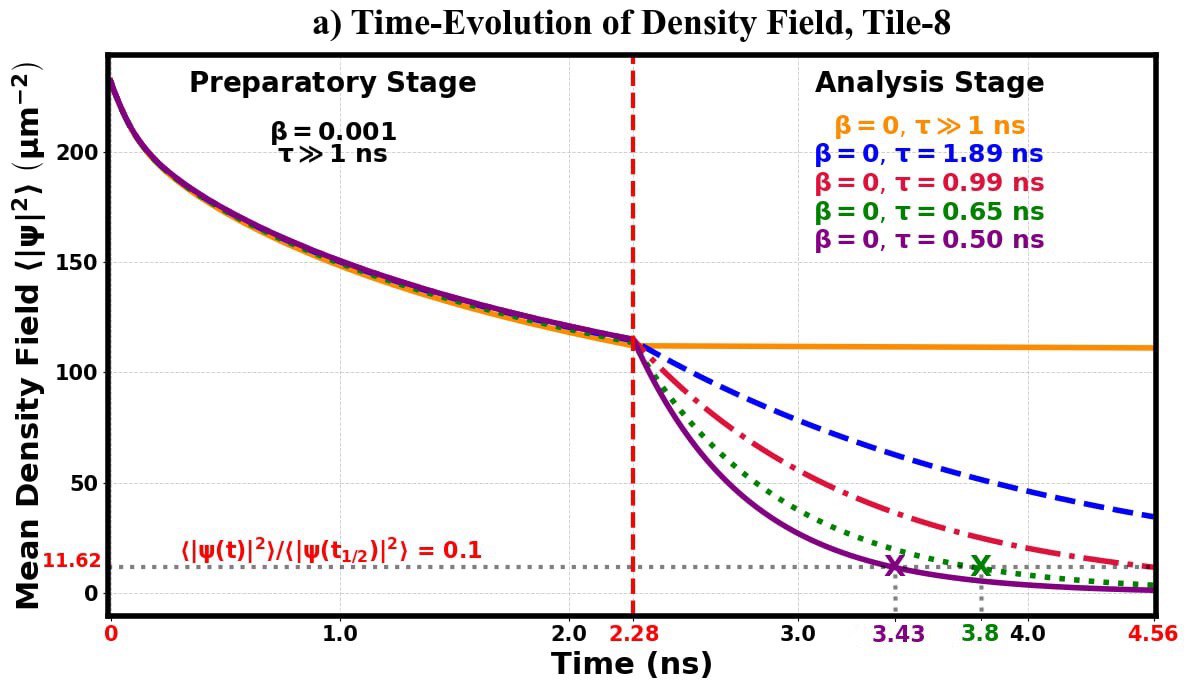}\\
        \includegraphics[width=\linewidth]{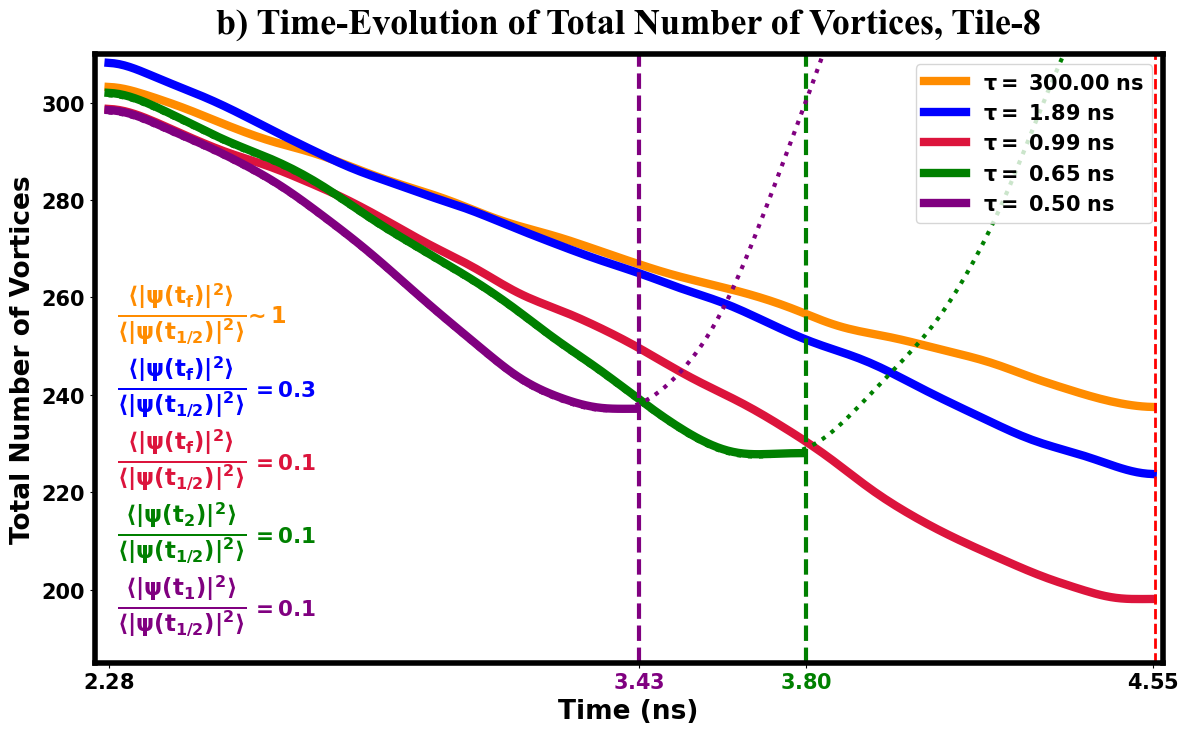}
    \end{center}
    
    \caption{\label{fig_5_decay_dens_and_Nvort} Time evolution of quantum fluid density \textbf{(a)} and total number of vortices \textbf{(b)} for various values of quantum fluid lifetime $\tau$ for Tile-8 excitation. The cross markers in panel \textbf{a)} correspond to the kink in number of vortices. \\  
    }
\end{figure}

In present section, we investigate at which extent quantum fluid density depletion effects on IKE spectra and vortex statistics of quantum fluid with losses (radiative losses in case of polaritons). At this stage of the analysis, it has been clear that Tile-8 imprinting is a valid quantum fluid excitation method to study quantum turbulence, as shown by the analysis of the IKE spectra and vortex statistics. So, we focus on this excitation method. First, we track the time evolution of quantum fluid density and total number of vortices, see Fig.~\ref{fig_5_decay_dens_and_Nvort} panels a) and b), respectively. One sees that for high lifetime the decay of the total number of vortices is very close to the conservative case. For stronger decay, the decrease of the total number of vortices is stronger. When the density reaches approximately 20~$\mu \mathrm{m}^{-2}$ (or 10\% of density at the beginning of the analysis phase), an abrupt increase in the number of vortices in the automatized detection procedure becomes visible, which corresponds to turbulence vanishment associated with density depletion and appearance of the noise in the phase field, see Appendix Fig. S6. However, for lifetimes higher than 1~ps, turbulence exists for the whole observation period.

\begin{figure}
    \begin{center}
        \includegraphics[width=\linewidth]{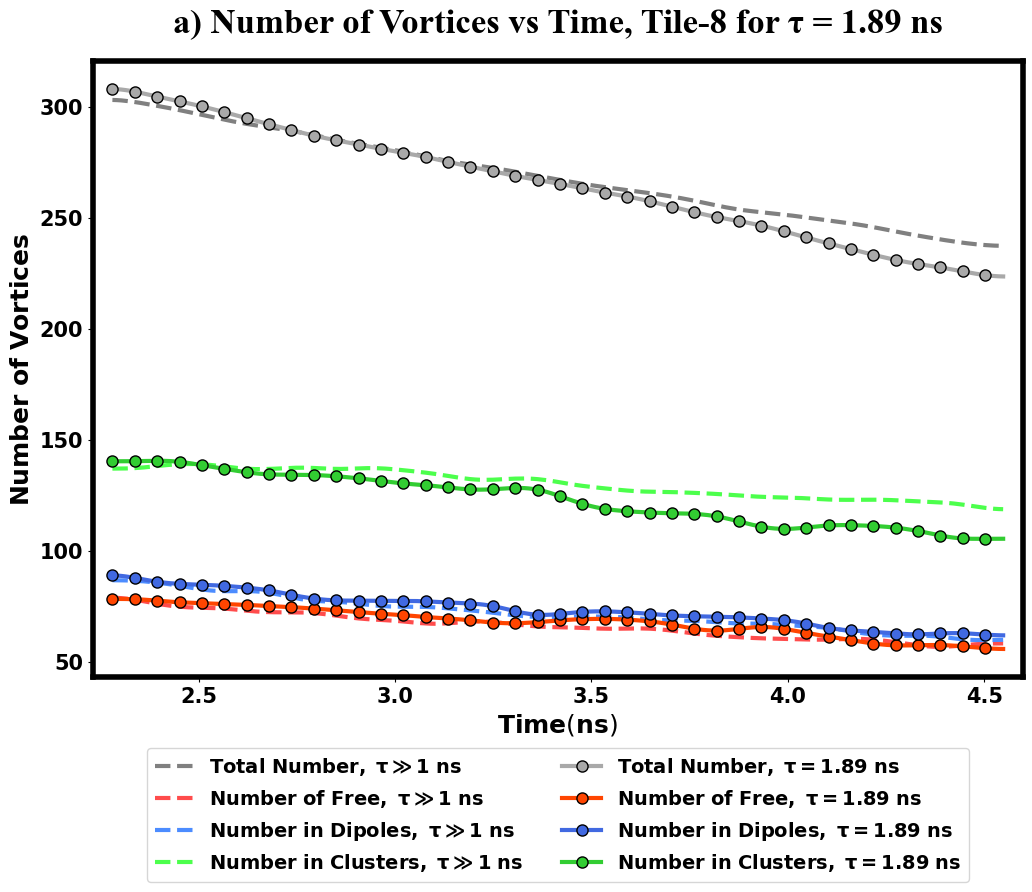}\\
        \includegraphics[width=\linewidth]{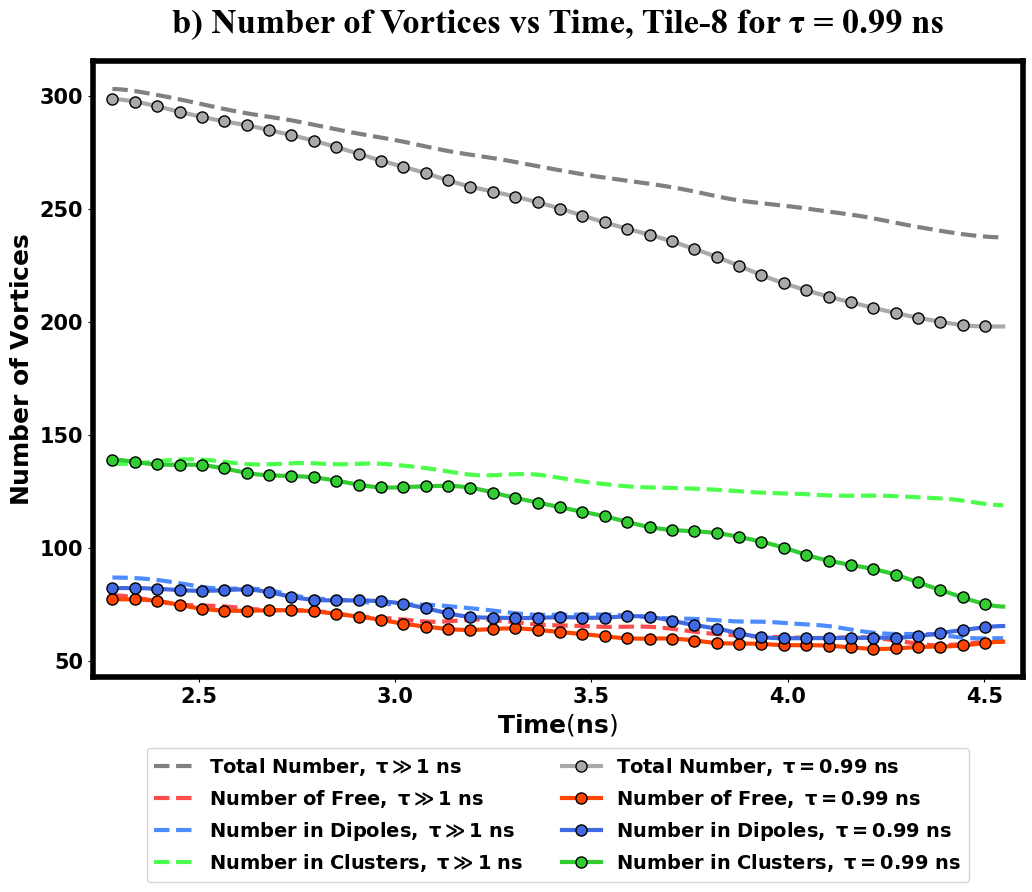}
    \end{center}
    
    \caption{\label{fig_6_decay_finer_statistics} Time evolution of the number of single vortices (red color), vortices in dipoles (blue) and in clusters (green) for lifetimes $\tau=1.89$~ns (panel \textbf{a}) and for lifetime $\tau=0.99$~ns (panel \textbf{b}). Gray color is for total amount of vortices. Dashed curves give the results for the conservative case.
    }
\end{figure}

Fig. \ref{fig_6_decay_finer_statistics} shows separation of the whole vortex ensemble into single vortices, dipoles and clusters for $\tau=1.89$~ns and $\tau=0.99$~ns compared to the conservative case. One can see that the lifetime for the typical polariton parameters (effective mass and interaction constant) giving the vortex evaporation dynamics as in the conservative case should be higher than 1~ns. Direct comparison of the vortex statistics after 2.38~ns of evolution with the initial one is given above in Fig.~\ref{fig_3_hist}. It is worth mentioning that the chosen parameters except lifetimes are typical for the exciton-polaritons in semiconductor microstructures and therefore the conclusions regarding the required lifetimes are quite general for the exciton-polariton systems.

\begin{figure}[t]
    \begin{center}
        \includegraphics[width=\linewidth]{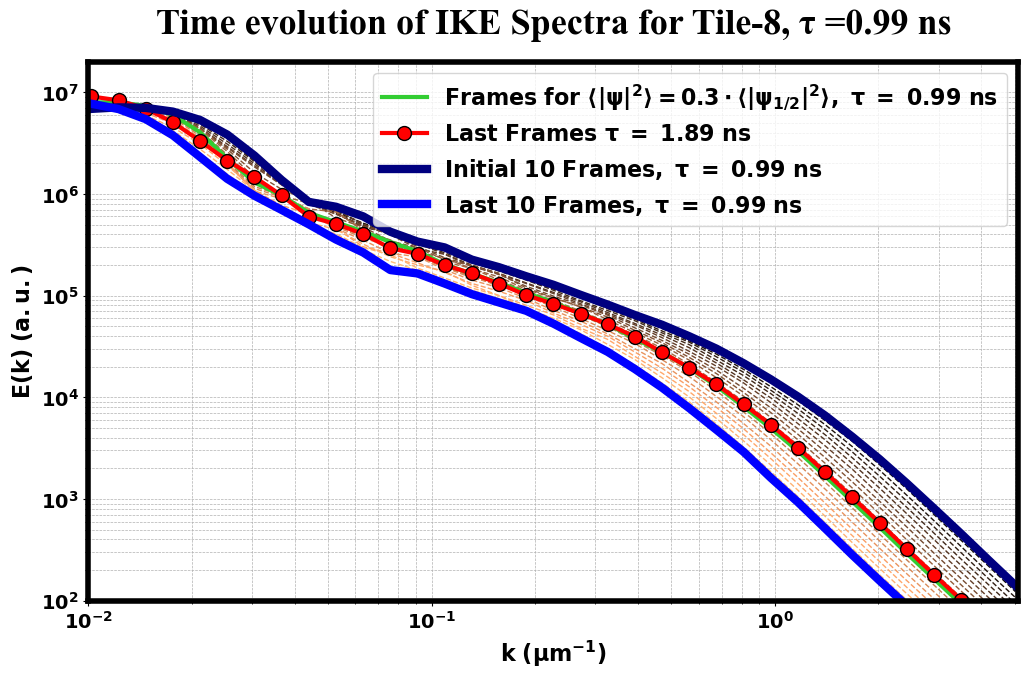}
    \end{center}
    \caption{\label{fig_7_IKE_decay} IKE spectra evolution for decaying quantum fluid with lifetime $\tau=1.89$~ns. Dark blue curve is for initial IKE spectrum and blue one is for spectrum after 2.38~ns evolution. Dashed lines in blue-orange gradient colors show intermediate stages for each 114~ps. Green curve shows spectrum for time moment 3.57~ns where the density is 0.3 of initial density. Red curve is given for comparison with the case of differing lifetime ($\tau=0.99$~ns) and density also of 0.3 of initial one at the final stage of simulation (4.56~ns). One can conclude that IKE spectrum depends primarily on the density but not on the decay rate if decay is slow enough.
    }
\end{figure}

It is also important to analyze the IKE spectra during the quantum fluid decay process. Fig.~\ref{fig_7_IKE_decay} shows the semi-analytical IKE spectra of the quantum fluid with finite lifetime $\tau = 1.89$~ns at various time moments. One can see the smooth evolution of spectra with time visible in overall lowering of incompressible kinetic energy at all time scales. At the same time, decay at wave vectors of the order of $\xi^{-1}$ and higher is stronger than that at low wave vectors, which is a fingerprint of the conservation of the global rotational motion.

\begin{figure}
    \begin{center}
        \includegraphics[width=\linewidth]{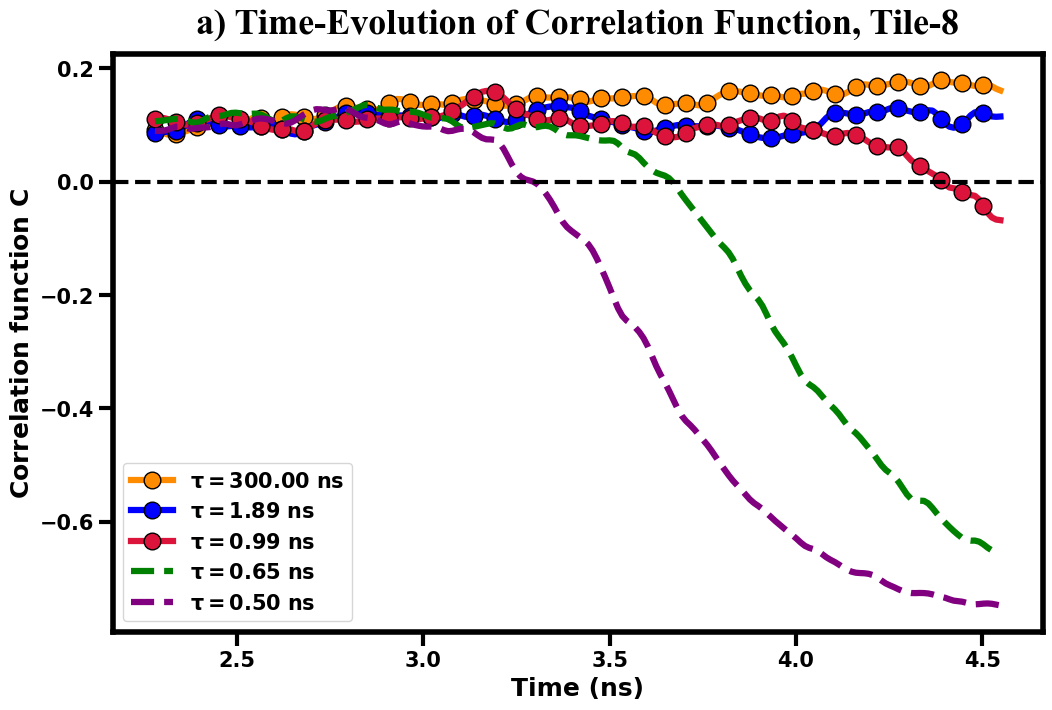}\\
        \includegraphics[width=\linewidth]{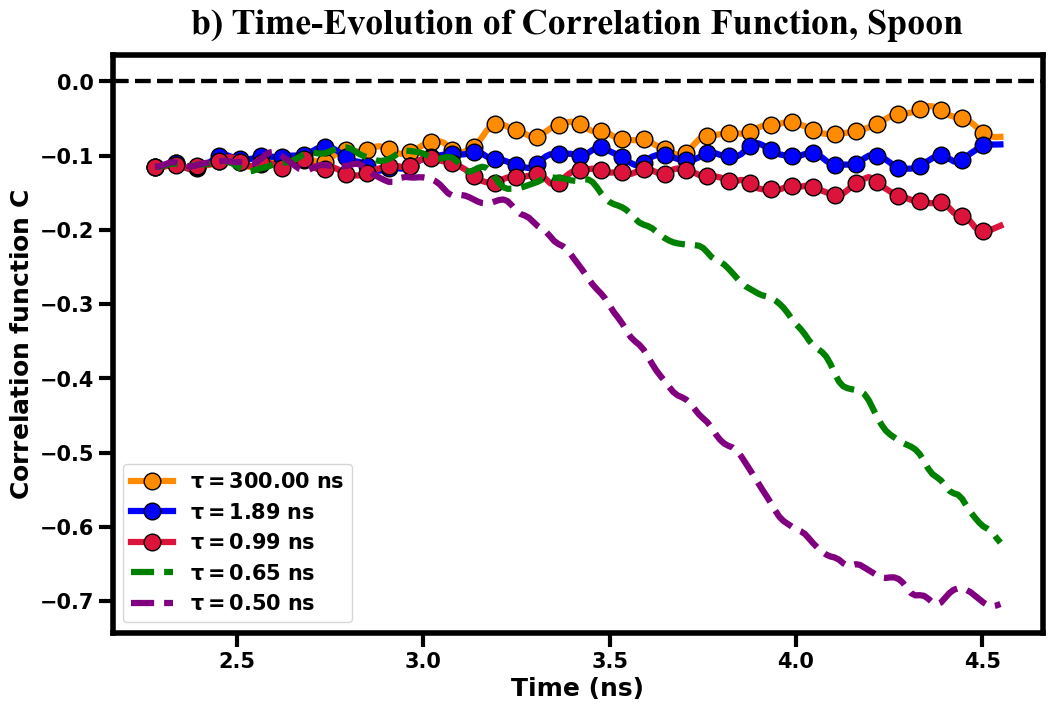}
    \end{center}
    
    \caption{\label{fig_8_decay_correaltion_function} Time evolution of the nearest-neighbor vorticity sign correlation function $C$ for various lifetimes for Tile-8 imprinting (panel \textbf{a}) and spoon stirring (panel \textbf{b}) strategies. Dashed curves correspond to the noise-like state of the low-density quantum fluid and the turbulence vanished.} 
\end{figure}

Moreover, we are interested in further analyzing (using the tile-imprinting scheme) formation of clouds of vortices with the same sign as polariton decay, which can be quantified via the nearest-neighbor sign correlation function $C$ as in Ref. \cite{panico2021dynamics}. In fact, non-zero positive values of the function $C$ highlight the emergence of a spontaneous order from chaotic motion, resulting in a low entropic state (negative vortex temperature) ~\cite{kraichnan1967inertial, simula2014emergence, groszek2018vortex}. 
Fig. \ref{fig_8_decay_correaltion_function} shows, for Tile-8 (panel a) and classical spoon (panel b) strategies, the time evolution of correlation function $C$ evaluated at the previously considered different quantum fluid loss rates (or equivalently polariton lifetimes). The behavior of the correlation function clearly confirms the existence of clouds of vortices with the same circulation sign for Tile-8 during all the entire simulation duration for $\tau=1.89, \ 0.99$~ns, as well as decaying and vanishing turbulence with a fast transition towards the negative values for lifetime values below 1~ns. Interestingly, that unlike ambiguous results of vortex statistics and IKE spectra analysis, correlation function is capable to distinguish Tile-8 and spoon stirring excitation showing higher organization in the former. One can notice how for Tile-8 excitation strategy more pronounced vortex clustering originates from the formation of ``streets of vortices'' at tile borders since the very initial evolution stage, see Fig.\ref{fig_2_detection}(d).

For both Tile-8 and spoon-stirring approaches, one can see the slow growth of $C$ correlation function corresponding to the evaporative-heating mechanism arising from the vortex annihilation occurring at the border of trap and in vortex dipoles, in agreement with Ref.~\cite{comaron2024dynamics}. From the given analysis, one can conclude that it is necessary to maintain the turbulence for at least several nanoseconds (and thus have the comparable lifetime) to detect the evaporative-heating effect fingerprints. A dissipative phenomenon as the radiative polariton decay significantly compromises this process, because of vortex core expansion slowing down the vortex dynamics, with a consequent conversion of IKE into a compressible energy and emergence of a several random distributed vortex singularities in the system. The important outcome is that for slow decay, IKE spectrum depends on initial configuration and on quantum fluids density, but not on the decay itself.

\subsection{Quantum turbulence in the regimes of fluctuating gain and loss}
\label{subsecIIIc_gain/loss_fluct}

In the present section, we investigate the possibility of maintaining quantum turbulence in the regime with alternating phases of positive and negative gain. For polaritons, this regime approximates the system behavior under incoherent laser pump with time-dependent power operating above and below the threshold. We intend to characterize the effects of gain/loss fluctuations, possibly arising either from intentionally modulated lasers or from experimental limitations on the optical laser stability.

\begin{figure}
    \begin{center}
        \includegraphics[width=\linewidth]{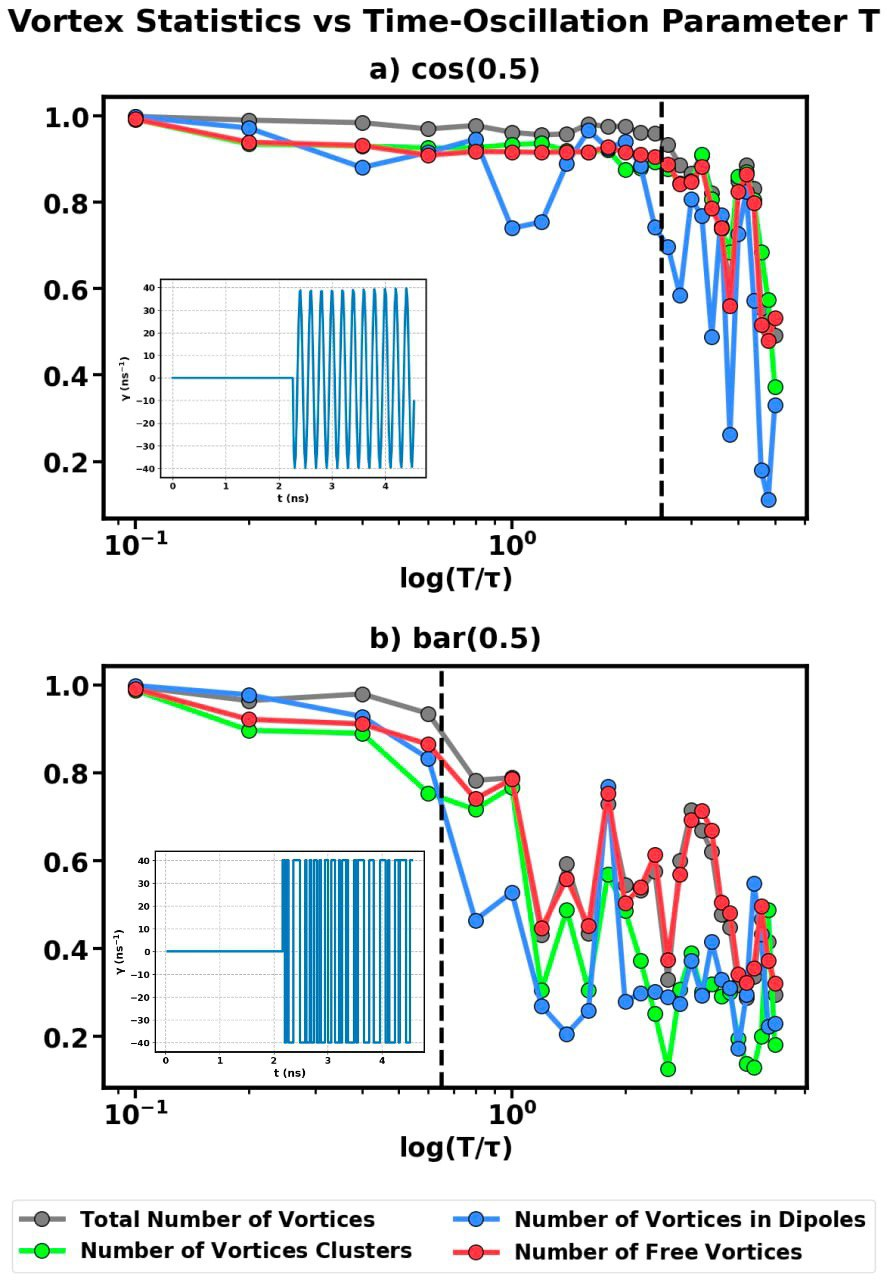}\\
    \end{center}
    
    \caption{\label{fig_9_vortex-statistics_cos05_bar05} Sustainability of quantum turbulence for the regime of alternating gain/loss with characteristic modulation period $T$. The quantity $J(X(T)) = \mathrm{min}(X(T), X_{\mathrm{conserv}})/\mathrm{max}(X(T), X_{\mathrm{conserv}})$ is plotted for $X$ standing for vortex statistical properties: number of all vortices, single vortices, vortices in clusters and dipoles at the end of the simulation. Panel \textbf{a)} is for cosine modulation and panel \textbf{b)} is for bar-shaped modulation. The right side from the vertical dashed lines refers to the regimes where the turbulence is significantly affected by gain/loss modulation. The insets demonstrate typical $\gamma(t)$ profiles.}
\end{figure}

We consider two regimes of gain/loss modulation. For the ``cosine'' regime, we set $\gamma(t) = \gamma_0\cdot \cos(2\pi/T \cdot t)$ in Eq.~\ref{eq: modified GPE}, where effective $\gamma_0^{-1} = \tau = 25$~ps, which is close to the values of lifetime in the actual polariton platforms. Practically, $\gamma(t)$ is defined by the combination of polariton radiative losses and stimulated scattering from the exciton reservoir. The ``bar'' regime is designed to be more stochastic. It corresponds to a rectangular profile of absolute value $\gamma_0$ modulated with alternating positive and negative signs with characteristic duration of the order of $T$, to be specific taken as a random value from the range $[0.1,\dots,1]\cdot T$. The insets in Fig.~\ref{fig_9_vortex-statistics_cos05_bar05} illustrate the employed modulation schemes. To quantify the sustainability of the turbulence, we employ the following procedure. First, we repeat the vortex detection procedure described in the previous subsections. Then we calculate the value $J(X(T)) = \mathrm{min}(X(T), X_{\mathrm{conserv}})/\mathrm{max}(X(T), X_{\mathrm{conserv}})$ that can be referred to as the ``Jaccard similarity for the values'', where $X(T)$ stands for the number of all vortices, vortices in clusters and dipoles, and single vortices at the end of the simulation for a given characteristic modulation period $T$. For significant difference with conservative case, i.e. $X(T)\gg X_{\mathrm{conserv}}$ or $X(T)\ll X_{\mathrm{conserv}}$ one obtains $J$ values are close to 0. The main panels in Fig.~\ref{fig_9_vortex-statistics_cos05_bar05} demonstrate $J$ values for cosine and bar types of gain/loss modulation. One sees that the turbulence remains maintained at $T/\tau$ of the order of unity, again demonstrating the condition of keeping fluctuations of the quantum fluid density less than one order in magnitude. For cosine modulation, the turbulence is maintained at a higher level $T$ due to the absence of long loss periods.

\begin{figure}
    \begin{center}
        \includegraphics[width=\linewidth]{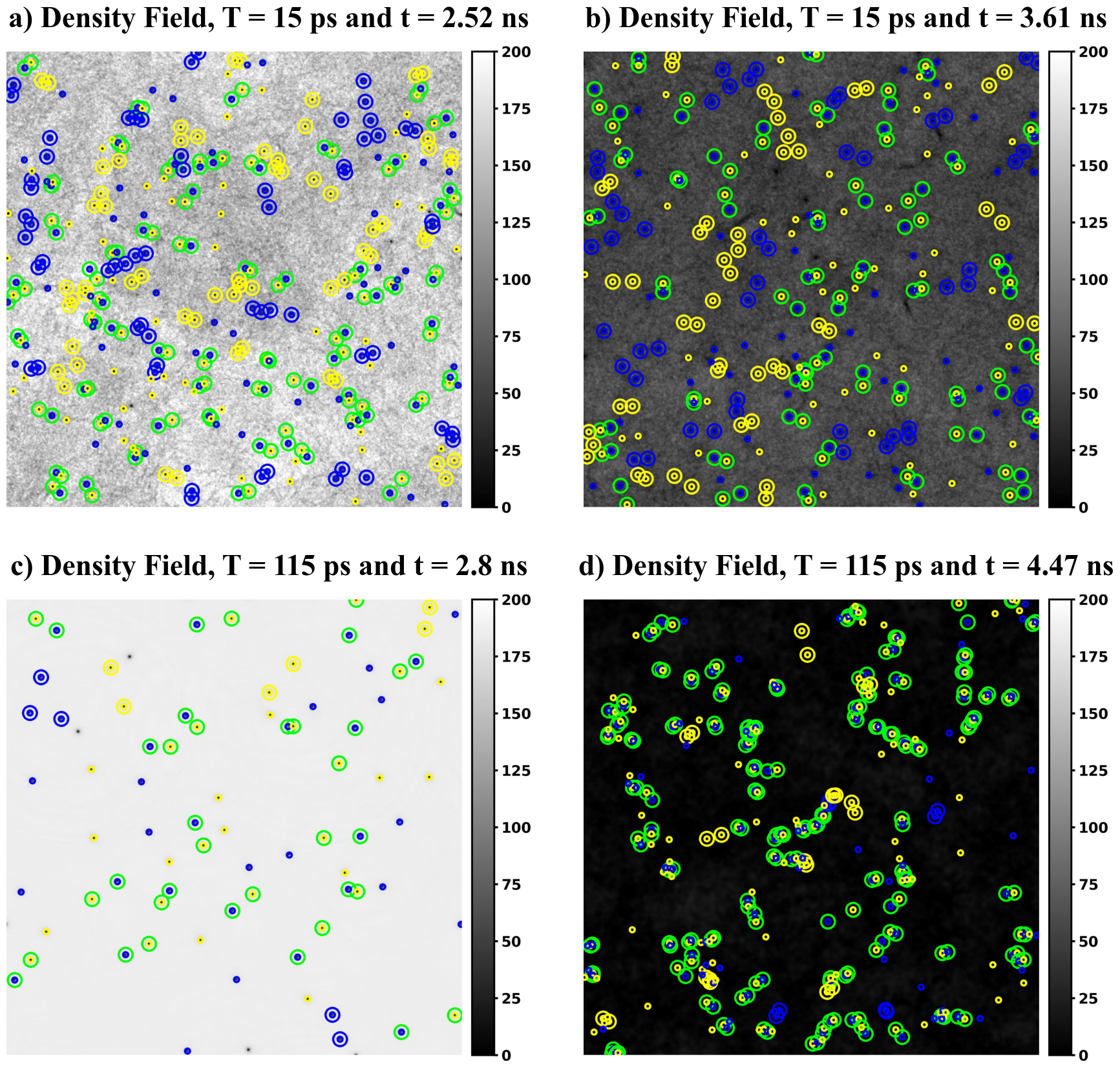}\\
    \end{center}
    
    \caption{\label{fig_10_bar05_density-snapshohts}\
    Results of vortex detection followed by clustering algorithm for bar-shaped time modulation with $T=15$~ps $= 0.6\cdot\tau$ \textbf{(a, b)} and $T=115$~ps $= 4.6\cdot\tau$ \textbf{(c, d)}. The panels on the left (\textbf{a, c}) show snapshot at the beginning of the evolution. Instead, on the right panels \textbf{(b, d)} show frames for the dissipation phase.}
\end{figure}

This is also visible in Fig.\ref{fig_10_bar05_density-snapshohts}, which shows the results of vortex detection and the application of the clustering algorithm for the cases of $T/\tau = 0.6$ at initial stages of evolution and during the loss-dominating interval and for higher modulation period $T/\tau = 4.6$. In the first case, turbulent motion is still reliable, due to smaller oscillations of the mean density, around the initial density $n_0=|\psi_0|^2=200 ~\mu$m$^2$, differently from the $T/\tau=4.6$ cases, in which during the depletion intervals the strong dissipation generates new randomly distributed vortex pairs.

Based on these evidences, one can conclude that the quantum turbulence evolution remains unperturbed only if the time scale of gain/loss modulation is smaller than the magnitude of modulation or at most comparable to it. Thus, potential utility of the polariton platform for quantum turbulence studies can be connected with the experimental protocol of the laser periodically operating above and below the condensation threshold. The positive gain phase would allow for the overcoming of polariton losses, while the negative gain phase would provide the unperturbed evolution to turbulent motion of the condensate wave function.

\section{Conclusion}
\label{secIV_conslusion}
In this work, we studied the 2D quantum turbulence with a focus on the inclusion of important specialties inherent to quantum fluids of exciton-polaritons in semiconductor microcavities. The first exciton-polariton feature complication the turbulence studies in a non-trivial implementation of the spoon-stirring vortex injection for turbulence development. However, we demonstrate robustly that one can effectively generate turbulence by direct imprinting the wave function with homogeneous magnitude and phase formed of tiles with randomly oriented plane waves. In experimental implementation of this scheme, one can use SLM with quasi-resonant laser excitation. The size of the tiles should be comparable to the desired distance between the layers. The created pattern automatically creates phase singularities, which then evolve into quantized vortices. After a certain free evolution period in the preparatory phase, the resulting statistics of the vortex and incompressible kinetic energy spectra are very similar to those of the conventional spoon-stirring technique. For clustered vortex only, we see the -5/3 exponent in the IKE spectrum that is related to Kolmogorov scaling law and to the fractal structures of clustered vortices. In addition to clear -5/3 exponent above the forcing scale~\cite{muller2024exploring}, existence of the associated inverse energy cascade was evidenced by the explicit tracing the time-dynamics of energy distribution~\cite{Koniakhin2020}. The slightly higher clusterization for Tile-8 scheme is visible in the $C$ correlation function for the signs of nearest neighbors. The randomness of the imprinted wave vector directions in Tile-8 provides the inherent stochasticity of quantum turbulence and is useful for averaging both for numerical simulations and for experimental realizations. At the same time, the imprinting scheme can allow one to collect the signal in repetitive experiments. Therefore, we conclude that the Tile-8 imprinting scheme is a valid alternative to conventional spoon stirring to study quantum turbulence in quantum fluids. Excitation with pixel random phase is also capable to generate vortices, however, efficiency of generation is significantly lower in this case. Moreover, creation of turbulent-like flows and clusters is also less effective for pixel schemes, which means that the tile schemes posses important features not occurring in more random approach. It is worth mentioning that, for the condensates in a parabolic traps, the spectral representation of the wave function can be employed for the effective solution of GPE~\cite{berman2015spontaneous}. In that case, it is natural to perform an analysis of the mode occupation number revealing the distribution reminiscent of the wave-turbulent state for the dipolar exciton condensate~\cite{berman2012turbulence}. However, for strong turbulence with multiple quantized vortices, the chosen numerical scheme is preferable as well as the metrics under analysis: vortex statistics and IKE spectra.  In terms of rotational motion scale, the actual turbulent regime lies between that less chaotic wave turbulence-like regime and the regime of Einstein-Bose condensed vortices~\cite{valani2018einstein}.

Secondly, we show that quantum turbulence can be robust with respect to polariton decay. The statistical quantities like the total number of vortices, the fraction of single and clustered vortices, fraction of vortices belonging to the dipoles evolve smoothly with depletion of the polariton density. The spectral characteristics of the quantum fluid also evolve smoothly. For sufficiently slow decay rate (high lifetimes), IKE spectra do not depend on the decay rate itself, but only on the current density. In general, the fingerprints of turbulence in statistics and IKE spectra remain until approximately one-order decrease of the quantum fluid density. With respect to observation of the evaporative heating effect, one should track the system for at least several nanoseconds for the typical polariton mass and interaction strength, which imposes the respective limitation on the lifetime. In case of enough long-living polaritions comparable to that in Refs.~\cite{nelsen2013dissipationless,panico2023onset} or higher, no need in the additional incoherent drive will be needed, thus excluding the exciton reservoir effects. Finally, we justify that gain/loss fluctuations do not affect the observation of turbulence under certain conditions on the involved time scales, highlighting the potential utility in quantum turbulence experiments of an experimental scheme of the laser periodically operating above and below the condensation threshold with characteristic period less than polariton lifetime.

\section*{Acknowledgements} The authors acknowledge the financial support from the Institute for Basic Science (IBS) in the Republic of Korea through the YSF project IBS-R024-Y3. We thank Oleg I. Utesov for useful remarks.

\bibliography{bib}

\begin{thebibliography}{43}%
\makeatletter
\providecommand \@ifxundefined [1]{%
 \@ifx{#1\undefined}
}%
\providecommand \@ifnum [1]{%
 \ifnum #1\expandafter \@firstoftwo
 \else \expandafter \@secondoftwo
 \fi
}%
\providecommand \@ifx [1]{%
 \ifx #1\expandafter \@firstoftwo
 \else \expandafter \@secondoftwo
 \fi
}%
\providecommand \natexlab [1]{#1}%
\providecommand \enquote  [1]{``#1''}%
\providecommand \bibnamefont  [1]{#1}%
\providecommand \bibfnamefont [1]{#1}%
\providecommand \citenamefont [1]{#1}%
\providecommand \href@noop [0]{\@secondoftwo}%
\providecommand \href [0]{\begingroup \@sanitize@url \@href}%
\providecommand \@href[1]{\@@startlink{#1}\@@href}%
\providecommand \@@href[1]{\endgroup#1\@@endlink}%
\providecommand \@sanitize@url [0]{\catcode `\\12\catcode `\$12\catcode `\&12\catcode `\#12\catcode `\^12\catcode `\_12\catcode `\%12\relax}%
\providecommand \@@startlink[1]{}%
\providecommand \@@endlink[0]{}%
\providecommand \url  [0]{\begingroup\@sanitize@url \@url }%
\providecommand \@url [1]{\endgroup\@href {#1}{\urlprefix }}%
\providecommand \urlprefix  [0]{URL }%
\providecommand \Eprint [0]{\href }%
\providecommand \doibase [0]{https://doi.org/}%
\providecommand \selectlanguage [0]{\@gobble}%
\providecommand \bibinfo  [0]{\@secondoftwo}%
\providecommand \bibfield  [0]{\@secondoftwo}%
\providecommand \translation [1]{[#1]}%
\providecommand \BibitemOpen [0]{}%
\providecommand \bibitemStop [0]{}%
\providecommand \bibitemNoStop [0]{.\EOS\space}%
\providecommand \EOS [0]{\spacefactor3000\relax}%
\providecommand \BibitemShut  [1]{\csname bibitem#1\endcsname}%
\let\auto@bib@innerbib\@empty
\bibitem [{\citenamefont {Vinen}(1957)}]{vinen1957mutual}%
  \BibitemOpen
  \bibfield  {author} {\bibinfo {author} {\bibfnamefont {W.~F.}\ \bibnamefont {Vinen}},\ }\bibfield  {title} {\bibinfo {title} {Mutual friction in a heat current in liquid helium ii i. experiments on steady heat currents},\ }\href@noop {} {\bibfield  {journal} {\bibinfo  {journal} {Proceedings of the Royal Society of London. Series A. Mathematical and Physical Sciences}\ }\textbf {\bibinfo {volume} {240}},\ \bibinfo {pages} {114} (\bibinfo {year} {1957})}\BibitemShut {NoStop}%
\bibitem [{\citenamefont {Skrbek}(2011)}]{skrbek2011quantum}%
  \BibitemOpen
  \bibfield  {author} {\bibinfo {author} {\bibfnamefont {L.}~\bibnamefont {Skrbek}},\ }\bibfield  {title} {\bibinfo {title} {Quantum turbulence},\ }in\ \href@noop {} {\emph {\bibinfo {booktitle} {Journal of Physics: Conference Series}}},\ Vol.\ \bibinfo {volume} {318}\ (\bibinfo {organization} {IOP Publishing},\ \bibinfo {year} {2011})\ p.\ \bibinfo {pages} {012004}\BibitemShut {NoStop}%
\bibitem [{\citenamefont {Tsubota}\ \emph {et~al.}(2013)\citenamefont {Tsubota}, \citenamefont {Kobayashi},\ and\ \citenamefont {Takeuchi}}]{tsubota2013quantum}%
  \BibitemOpen
  \bibfield  {author} {\bibinfo {author} {\bibfnamefont {M.}~\bibnamefont {Tsubota}}, \bibinfo {author} {\bibfnamefont {M.}~\bibnamefont {Kobayashi}},\ and\ \bibinfo {author} {\bibfnamefont {H.}~\bibnamefont {Takeuchi}},\ }\bibfield  {title} {\bibinfo {title} {Quantum hydrodynamics},\ }\href@noop {} {\bibfield  {journal} {\bibinfo  {journal} {Physics Reports}\ }\textbf {\bibinfo {volume} {522}},\ \bibinfo {pages} {191} (\bibinfo {year} {2013})}\BibitemShut {NoStop}%
\bibitem [{\citenamefont {White}\ \emph {et~al.}(2014)\citenamefont {White}, \citenamefont {Anderson},\ and\ \citenamefont {Bagnato}}]{white2014vortices}%
  \BibitemOpen
  \bibfield  {author} {\bibinfo {author} {\bibfnamefont {A.~C.}\ \bibnamefont {White}}, \bibinfo {author} {\bibfnamefont {B.~P.}\ \bibnamefont {Anderson}},\ and\ \bibinfo {author} {\bibfnamefont {V.~S.}\ \bibnamefont {Bagnato}},\ }\bibfield  {title} {\bibinfo {title} {Vortices and turbulence in trapped atomic condensates},\ }\href@noop {} {\bibfield  {journal} {\bibinfo  {journal} {Proceedings of the National Academy of Sciences}\ }\textbf {\bibinfo {volume} {111}},\ \bibinfo {pages} {4719} (\bibinfo {year} {2014})}\BibitemShut {NoStop}%
\bibitem [{\citenamefont {Tsubota}\ \emph {et~al.}(2017)\citenamefont {Tsubota}, \citenamefont {Fujimoto},\ and\ \citenamefont {Yui}}]{tsubota2017numerical}%
  \BibitemOpen
  \bibfield  {author} {\bibinfo {author} {\bibfnamefont {M.}~\bibnamefont {Tsubota}}, \bibinfo {author} {\bibfnamefont {K.}~\bibnamefont {Fujimoto}},\ and\ \bibinfo {author} {\bibfnamefont {S.}~\bibnamefont {Yui}},\ }\bibfield  {title} {\bibinfo {title} {Numerical studies of quantum turbulence},\ }\href@noop {} {\bibfield  {journal} {\bibinfo  {journal} {Journal of Low Temperature Physics}\ }\textbf {\bibinfo {volume} {188}},\ \bibinfo {pages} {119} (\bibinfo {year} {2017})}\BibitemShut {NoStop}%
\bibitem [{\citenamefont {Simula}\ \emph {et~al.}(2014)\citenamefont {Simula}, \citenamefont {Davis},\ and\ \citenamefont {Helmerson}}]{simula2014emergence}%
  \BibitemOpen
  \bibfield  {author} {\bibinfo {author} {\bibfnamefont {T.}~\bibnamefont {Simula}}, \bibinfo {author} {\bibfnamefont {M.~J.}\ \bibnamefont {Davis}},\ and\ \bibinfo {author} {\bibfnamefont {K.}~\bibnamefont {Helmerson}},\ }\bibfield  {title} {\bibinfo {title} {Emergence of order from turbulence in an isolated planar superfluid},\ }\href@noop {} {\bibfield  {journal} {\bibinfo  {journal} {Physical review letters}\ }\textbf {\bibinfo {volume} {113}},\ \bibinfo {pages} {165302} (\bibinfo {year} {2014})}\BibitemShut {NoStop}%
\bibitem [{\citenamefont {Valani}\ \emph {et~al.}(2018)\citenamefont {Valani}, \citenamefont {Groszek},\ and\ \citenamefont {Simula}}]{valani2018einstein}%
  \BibitemOpen
  \bibfield  {author} {\bibinfo {author} {\bibfnamefont {R.~N.}\ \bibnamefont {Valani}}, \bibinfo {author} {\bibfnamefont {A.~J.}\ \bibnamefont {Groszek}},\ and\ \bibinfo {author} {\bibfnamefont {T.~P.}\ \bibnamefont {Simula}},\ }\bibfield  {title} {\bibinfo {title} {{E}instein--{B}ose condensation of {O}nsager vortices},\ }\href@noop {} {\bibfield  {journal} {\bibinfo  {journal} {New Journal of Physics}\ }\textbf {\bibinfo {volume} {20}},\ \bibinfo {pages} {053038} (\bibinfo {year} {2018})}\BibitemShut {NoStop}%
\bibitem [{\citenamefont {Horng}\ \emph {et~al.}(2009)\citenamefont {Horng}, \citenamefont {Hsueh}, \citenamefont {Su}, \citenamefont {Kao},\ and\ \citenamefont {Gou}}]{horng2009two}%
  \BibitemOpen
  \bibfield  {author} {\bibinfo {author} {\bibfnamefont {T.-L.}\ \bibnamefont {Horng}}, \bibinfo {author} {\bibfnamefont {C.-H.}\ \bibnamefont {Hsueh}}, \bibinfo {author} {\bibfnamefont {S.-W.}\ \bibnamefont {Su}}, \bibinfo {author} {\bibfnamefont {Y.-M.}\ \bibnamefont {Kao}},\ and\ \bibinfo {author} {\bibfnamefont {S.-C.}\ \bibnamefont {Gou}},\ }\bibfield  {title} {\bibinfo {title} {Two-dimensional quantum turbulence in a nonuniform {B}ose-{E}instein condensate},\ }\href@noop {} {\bibfield  {journal} {\bibinfo  {journal} {Physical Review A—Atomic, Molecular, and Optical Physics}\ }\textbf {\bibinfo {volume} {80}},\ \bibinfo {pages} {023618} (\bibinfo {year} {2009})}\BibitemShut {NoStop}%
\bibitem [{\citenamefont {Billam}\ \emph {et~al.}(2015)\citenamefont {Billam}, \citenamefont {Reeves},\ and\ \citenamefont {Bradley}}]{billam2015spectral}%
  \BibitemOpen
  \bibfield  {author} {\bibinfo {author} {\bibfnamefont {T.}~\bibnamefont {Billam}}, \bibinfo {author} {\bibfnamefont {M.}~\bibnamefont {Reeves}},\ and\ \bibinfo {author} {\bibfnamefont {A.}~\bibnamefont {Bradley}},\ }\bibfield  {title} {\bibinfo {title} {Spectral energy transport in two-dimensional quantum vortex dynamics},\ }\href@noop {} {\bibfield  {journal} {\bibinfo  {journal} {Physical Review A}\ }\textbf {\bibinfo {volume} {91}},\ \bibinfo {pages} {023615} (\bibinfo {year} {2015})}\BibitemShut {NoStop}%
\bibitem [{\citenamefont {Numasato}\ \emph {et~al.}(2010)\citenamefont {Numasato}, \citenamefont {Tsubota},\ and\ \citenamefont {L’vov}}]{numasato2010direct}%
  \BibitemOpen
  \bibfield  {author} {\bibinfo {author} {\bibfnamefont {R.}~\bibnamefont {Numasato}}, \bibinfo {author} {\bibfnamefont {M.}~\bibnamefont {Tsubota}},\ and\ \bibinfo {author} {\bibfnamefont {V.~S.}\ \bibnamefont {L’vov}},\ }\bibfield  {title} {\bibinfo {title} {Direct energy cascade in two-dimensional compressible quantum turbulence},\ }\href@noop {} {\bibfield  {journal} {\bibinfo  {journal} {Physical Review A—Atomic, Molecular, and Optical Physics}\ }\textbf {\bibinfo {volume} {81}},\ \bibinfo {pages} {063630} (\bibinfo {year} {2010})}\BibitemShut {NoStop}%
\bibitem [{\citenamefont {Berloff}(2010)}]{berloff2010turbulence}%
  \BibitemOpen
  \bibfield  {author} {\bibinfo {author} {\bibfnamefont {N.~G.}\ \bibnamefont {Berloff}},\ }\bibfield  {title} {\bibinfo {title} {Turbulence in exciton-polariton condensates},\ }\href@noop {} {\bibfield  {journal} {\bibinfo  {journal} {arXiv preprint arXiv:1010.5225}\ } (\bibinfo {year} {2010})}\BibitemShut {NoStop}%
\bibitem [{\citenamefont {Panico}\ \emph {et~al.}(2023)\citenamefont {Panico}, \citenamefont {Comaron}, \citenamefont {Matuszewski}, \citenamefont {Lanotte}, \citenamefont {Trypogeorgos}, \citenamefont {Gigli}, \citenamefont {Giorgi}, \citenamefont {Ardizzone}, \citenamefont {Sanvitto},\ and\ \citenamefont {Ballarini}}]{panico2023onset}%
  \BibitemOpen
  \bibfield  {author} {\bibinfo {author} {\bibfnamefont {R.}~\bibnamefont {Panico}}, \bibinfo {author} {\bibfnamefont {P.}~\bibnamefont {Comaron}}, \bibinfo {author} {\bibfnamefont {M.}~\bibnamefont {Matuszewski}}, \bibinfo {author} {\bibfnamefont {A.}~\bibnamefont {Lanotte}}, \bibinfo {author} {\bibfnamefont {D.}~\bibnamefont {Trypogeorgos}}, \bibinfo {author} {\bibfnamefont {G.}~\bibnamefont {Gigli}}, \bibinfo {author} {\bibfnamefont {M.~D.}\ \bibnamefont {Giorgi}}, \bibinfo {author} {\bibfnamefont {V.}~\bibnamefont {Ardizzone}}, \bibinfo {author} {\bibfnamefont {D.}~\bibnamefont {Sanvitto}},\ and\ \bibinfo {author} {\bibfnamefont {D.}~\bibnamefont {Ballarini}},\ }\bibfield  {title} {\bibinfo {title} {Onset of vortex clustering and inverse energy cascade in dissipative quantum fluids},\ }\href@noop {} {\bibfield  {journal} {\bibinfo  {journal} {Nature Photonics}\ }\textbf {\bibinfo {volume} {17}},\ \bibinfo {pages} {451} (\bibinfo {year} {2023})}\BibitemShut {NoStop}%
\bibitem [{\citenamefont {Carusotto}\ and\ \citenamefont {Ciuti}(2004)}]{carusotto2004probing}%
  \BibitemOpen
  \bibfield  {author} {\bibinfo {author} {\bibfnamefont {I.}~\bibnamefont {Carusotto}}\ and\ \bibinfo {author} {\bibfnamefont {C.}~\bibnamefont {Ciuti}},\ }\bibfield  {title} {\bibinfo {title} {Probing microcavity polariton superfluidity through resonant rayleigh scattering},\ }\href@noop {} {\bibfield  {journal} {\bibinfo  {journal} {Physical review letters}\ }\textbf {\bibinfo {volume} {93}},\ \bibinfo {pages} {166401} (\bibinfo {year} {2004})}\BibitemShut {NoStop}%
\bibitem [{\citenamefont {Amo}\ \emph {et~al.}(2009)\citenamefont {Amo}, \citenamefont {Lefr{\`e}re}, \citenamefont {Pigeon}, \citenamefont {Adrados}, \citenamefont {Ciuti}, \citenamefont {Carusotto}, \citenamefont {Houdr{\'e}}, \citenamefont {Giacobino},\ and\ \citenamefont {Bramati}}]{amo2009superfluidity}%
  \BibitemOpen
  \bibfield  {author} {\bibinfo {author} {\bibfnamefont {A.}~\bibnamefont {Amo}}, \bibinfo {author} {\bibfnamefont {J.}~\bibnamefont {Lefr{\`e}re}}, \bibinfo {author} {\bibfnamefont {S.}~\bibnamefont {Pigeon}}, \bibinfo {author} {\bibfnamefont {C.}~\bibnamefont {Adrados}}, \bibinfo {author} {\bibfnamefont {C.}~\bibnamefont {Ciuti}}, \bibinfo {author} {\bibfnamefont {I.}~\bibnamefont {Carusotto}}, \bibinfo {author} {\bibfnamefont {R.}~\bibnamefont {Houdr{\'e}}}, \bibinfo {author} {\bibfnamefont {E.}~\bibnamefont {Giacobino}},\ and\ \bibinfo {author} {\bibfnamefont {A.}~\bibnamefont {Bramati}},\ }\bibfield  {title} {\bibinfo {title} {Superfluidity of polaritons in semiconductor microcavities},\ }\href@noop {} {\bibfield  {journal} {\bibinfo  {journal} {Nature Physics}\ }\textbf {\bibinfo {volume} {5}},\ \bibinfo {pages} {805} (\bibinfo {year} {2009})}\BibitemShut {NoStop}%
\bibitem [{\citenamefont {Kasprzak}\ \emph {et~al.}(2006)\citenamefont {Kasprzak}, \citenamefont {Richard}, \citenamefont {Kundermann}, \citenamefont {Baas}, \citenamefont {Jeambrun}, \citenamefont {Keeling}, \citenamefont {Marchetti}, \citenamefont {Szyma{\'n}ska}, \citenamefont {Andr{\'e}}, \citenamefont {Staehli} \emph {et~al.}}]{kasprzak2006bose}%
  \BibitemOpen
  \bibfield  {author} {\bibinfo {author} {\bibfnamefont {J.}~\bibnamefont {Kasprzak}}, \bibinfo {author} {\bibfnamefont {M.}~\bibnamefont {Richard}}, \bibinfo {author} {\bibfnamefont {S.}~\bibnamefont {Kundermann}}, \bibinfo {author} {\bibfnamefont {A.}~\bibnamefont {Baas}}, \bibinfo {author} {\bibfnamefont {P.}~\bibnamefont {Jeambrun}}, \bibinfo {author} {\bibfnamefont {J.~M.~J.}\ \bibnamefont {Keeling}}, \bibinfo {author} {\bibfnamefont {F.}~\bibnamefont {Marchetti}}, \bibinfo {author} {\bibfnamefont {M.}~\bibnamefont {Szyma{\'n}ska}}, \bibinfo {author} {\bibfnamefont {R.}~\bibnamefont {Andr{\'e}}}, \bibinfo {author} {\bibfnamefont {J.}~\bibnamefont {Staehli}}, \emph {et~al.},\ }\bibfield  {title} {\bibinfo {title} {Bose--{E}instein condensation of exciton polaritons},\ }\href@noop {} {\bibfield  {journal} {\bibinfo  {journal} {Nature}\ }\textbf {\bibinfo {volume} {443}},\ \bibinfo {pages} {409} (\bibinfo {year} {2006})}\BibitemShut {NoStop}%
\bibitem [{\citenamefont {Roumpos}\ \emph {et~al.}(2011)\citenamefont {Roumpos}, \citenamefont {Fraser}, \citenamefont {L{\"o}ffler}, \citenamefont {H{\"o}fling}, \citenamefont {Forchel},\ and\ \citenamefont {Yamamoto}}]{roumpos2011single}%
  \BibitemOpen
  \bibfield  {author} {\bibinfo {author} {\bibfnamefont {G.}~\bibnamefont {Roumpos}}, \bibinfo {author} {\bibfnamefont {M.~D.}\ \bibnamefont {Fraser}}, \bibinfo {author} {\bibfnamefont {A.}~\bibnamefont {L{\"o}ffler}}, \bibinfo {author} {\bibfnamefont {S.}~\bibnamefont {H{\"o}fling}}, \bibinfo {author} {\bibfnamefont {A.}~\bibnamefont {Forchel}},\ and\ \bibinfo {author} {\bibfnamefont {Y.}~\bibnamefont {Yamamoto}},\ }\bibfield  {title} {\bibinfo {title} {Single vortex--antivortex pair in an exciton-polariton condensate},\ }\href@noop {} {\bibfield  {journal} {\bibinfo  {journal} {Nature Physics}\ }\textbf {\bibinfo {volume} {7}},\ \bibinfo {pages} {129} (\bibinfo {year} {2011})}\BibitemShut {NoStop}%
\bibitem [{\citenamefont {Lagoudakis}\ \emph {et~al.}(2008)\citenamefont {Lagoudakis}, \citenamefont {Wouters}, \citenamefont {Richard}, \citenamefont {Baas}, \citenamefont {Carusotto}, \citenamefont {Andr{\'e}}, \citenamefont {Dang},\ and\ \citenamefont {Deveaud-Pl{\'e}dran}}]{lagoudakis2008quantized}%
  \BibitemOpen
  \bibfield  {author} {\bibinfo {author} {\bibfnamefont {K.~G.}\ \bibnamefont {Lagoudakis}}, \bibinfo {author} {\bibfnamefont {M.}~\bibnamefont {Wouters}}, \bibinfo {author} {\bibfnamefont {M.}~\bibnamefont {Richard}}, \bibinfo {author} {\bibfnamefont {A.}~\bibnamefont {Baas}}, \bibinfo {author} {\bibfnamefont {I.}~\bibnamefont {Carusotto}}, \bibinfo {author} {\bibfnamefont {R.}~\bibnamefont {Andr{\'e}}}, \bibinfo {author} {\bibfnamefont {L.~S.}\ \bibnamefont {Dang}},\ and\ \bibinfo {author} {\bibfnamefont {B.}~\bibnamefont {Deveaud-Pl{\'e}dran}},\ }\bibfield  {title} {\bibinfo {title} {Quantized vortices in an exciton--polariton condensate},\ }\href@noop {} {\bibfield  {journal} {\bibinfo  {journal} {Nature physics}\ }\textbf {\bibinfo {volume} {4}},\ \bibinfo {pages} {706} (\bibinfo {year} {2008})}\BibitemShut {NoStop}%
\bibitem [{\citenamefont {Dominici}\ \emph {et~al.}(2015)\citenamefont {Dominici}, \citenamefont {Dagvadorj}, \citenamefont {Fellows}, \citenamefont {Ballarini}, \citenamefont {De~Giorgi}, \citenamefont {Marchetti}, \citenamefont {Piccirillo}, \citenamefont {Marrucci}, \citenamefont {Bramati}, \citenamefont {Gigli} \emph {et~al.}}]{dominici2015vortex}%
  \BibitemOpen
  \bibfield  {author} {\bibinfo {author} {\bibfnamefont {L.}~\bibnamefont {Dominici}}, \bibinfo {author} {\bibfnamefont {G.}~\bibnamefont {Dagvadorj}}, \bibinfo {author} {\bibfnamefont {J.~M.}\ \bibnamefont {Fellows}}, \bibinfo {author} {\bibfnamefont {D.}~\bibnamefont {Ballarini}}, \bibinfo {author} {\bibfnamefont {M.}~\bibnamefont {De~Giorgi}}, \bibinfo {author} {\bibfnamefont {F.~M.}\ \bibnamefont {Marchetti}}, \bibinfo {author} {\bibfnamefont {B.}~\bibnamefont {Piccirillo}}, \bibinfo {author} {\bibfnamefont {L.}~\bibnamefont {Marrucci}}, \bibinfo {author} {\bibfnamefont {A.}~\bibnamefont {Bramati}}, \bibinfo {author} {\bibfnamefont {G.}~\bibnamefont {Gigli}}, \emph {et~al.},\ }\bibfield  {title} {\bibinfo {title} {Vortex and half-vortex dynamics in a nonlinear spinor quantum fluid},\ }\href@noop {} {\bibfield  {journal} {\bibinfo  {journal} {Science advances}\ }\textbf {\bibinfo {volume} {1}},\ \bibinfo {pages} {e1500807} (\bibinfo {year} {2015})}\BibitemShut {NoStop}%
\bibitem [{\citenamefont {Lagoudakis}\ \emph {et~al.}(2011)\citenamefont {Lagoudakis}, \citenamefont {Manni}, \citenamefont {Pietka}, \citenamefont {Wouters}, \citenamefont {Liew}, \citenamefont {Savona}, \citenamefont {Kavokin}, \citenamefont {Andr{\'e}},\ and\ \citenamefont {Deveaud-Pl{\'e}dran}}]{lagoudakis2011probing}%
  \BibitemOpen
  \bibfield  {author} {\bibinfo {author} {\bibfnamefont {K.~G.}\ \bibnamefont {Lagoudakis}}, \bibinfo {author} {\bibfnamefont {F.}~\bibnamefont {Manni}}, \bibinfo {author} {\bibfnamefont {B.}~\bibnamefont {Pietka}}, \bibinfo {author} {\bibfnamefont {M.}~\bibnamefont {Wouters}}, \bibinfo {author} {\bibfnamefont {T.~C.~H.}\ \bibnamefont {Liew}}, \bibinfo {author} {\bibfnamefont {V.}~\bibnamefont {Savona}}, \bibinfo {author} {\bibfnamefont {A.~V.}\ \bibnamefont {Kavokin}}, \bibinfo {author} {\bibfnamefont {R.}~\bibnamefont {Andr{\'e}}},\ and\ \bibinfo {author} {\bibfnamefont {B.}~\bibnamefont {Deveaud-Pl{\'e}dran}},\ }\bibfield  {title} {\bibinfo {title} {Probing the dynamics of spontaneous quantum vortices in polariton superfluids},\ }\href@noop {} {\bibfield  {journal} {\bibinfo  {journal} {Physical review letters}\ }\textbf {\bibinfo {volume} {106}},\ \bibinfo {pages} {115301} (\bibinfo {year} {2011})}\BibitemShut {NoStop}%
\bibitem [{\citenamefont {Madison}\ \emph {et~al.}(2000)\citenamefont {Madison}, \citenamefont {Chevy}, \citenamefont {Wohlleben},\ and\ \citenamefont {Dalibard}}]{madison2000vortex}%
  \BibitemOpen
  \bibfield  {author} {\bibinfo {author} {\bibfnamefont {K.~W.}\ \bibnamefont {Madison}}, \bibinfo {author} {\bibfnamefont {F.}~\bibnamefont {Chevy}}, \bibinfo {author} {\bibfnamefont {W.}~\bibnamefont {Wohlleben}},\ and\ \bibinfo {author} {\bibfnamefont {J.}~\bibnamefont {Dalibard}},\ }\bibfield  {title} {\bibinfo {title} {Vortex formation in a stirred bose-einstein condensate},\ }\href@noop {} {\bibfield  {journal} {\bibinfo  {journal} {Physical review letters}\ }\textbf {\bibinfo {volume} {84}},\ \bibinfo {pages} {806} (\bibinfo {year} {2000})}\BibitemShut {NoStop}%
\bibitem [{\citenamefont {Reeves}\ \emph {et~al.}(2012)\citenamefont {Reeves}, \citenamefont {Anderson},\ and\ \citenamefont {Bradley}}]{reeves2012classical}%
  \BibitemOpen
  \bibfield  {author} {\bibinfo {author} {\bibfnamefont {M.}~\bibnamefont {Reeves}}, \bibinfo {author} {\bibfnamefont {B.}~\bibnamefont {Anderson}},\ and\ \bibinfo {author} {\bibfnamefont {A.}~\bibnamefont {Bradley}},\ }\bibfield  {title} {\bibinfo {title} {Classical and quantum regimes of two-dimensional turbulence in trapped {B}ose-{E}instein condensates},\ }\href@noop {} {\bibfield  {journal} {\bibinfo  {journal} {Physical Review A—Atomic, Molecular, and Optical Physics}\ }\textbf {\bibinfo {volume} {86}},\ \bibinfo {pages} {053621} (\bibinfo {year} {2012})}\BibitemShut {NoStop}%
\bibitem [{\citenamefont {Reeves}\ \emph {et~al.}(2022)\citenamefont {Reeves}, \citenamefont {Goddard-Lee}, \citenamefont {Gauthier}, \citenamefont {Stockdale}, \citenamefont {Salman}, \citenamefont {Edmonds}, \citenamefont {Yu}, \citenamefont {Bradley}, \citenamefont {Baker}, \citenamefont {Rubinsztein-Dunlop} \emph {et~al.}}]{reeves2022turbulent}%
  \BibitemOpen
  \bibfield  {author} {\bibinfo {author} {\bibfnamefont {M.~T.}\ \bibnamefont {Reeves}}, \bibinfo {author} {\bibfnamefont {K.}~\bibnamefont {Goddard-Lee}}, \bibinfo {author} {\bibfnamefont {G.}~\bibnamefont {Gauthier}}, \bibinfo {author} {\bibfnamefont {O.~R.}\ \bibnamefont {Stockdale}}, \bibinfo {author} {\bibfnamefont {H.}~\bibnamefont {Salman}}, \bibinfo {author} {\bibfnamefont {T.}~\bibnamefont {Edmonds}}, \bibinfo {author} {\bibfnamefont {X.}~\bibnamefont {Yu}}, \bibinfo {author} {\bibfnamefont {A.~S.}\ \bibnamefont {Bradley}}, \bibinfo {author} {\bibfnamefont {M.}~\bibnamefont {Baker}}, \bibinfo {author} {\bibfnamefont {H.}~\bibnamefont {Rubinsztein-Dunlop}}, \emph {et~al.},\ }\bibfield  {title} {\bibinfo {title} {Turbulent relaxation to equilibrium in a two-dimensional quantum vortex gas},\ }\href@noop {} {\bibfield  {journal} {\bibinfo  {journal} {Physical Review X}\ }\textbf {\bibinfo {volume} {12}},\ \bibinfo {pages} {011031} (\bibinfo {year} {2022})}\BibitemShut {NoStop}%
\bibitem [{\citenamefont {Gauthier}\ \emph {et~al.}(2019)\citenamefont {Gauthier}, \citenamefont {Reeves}, \citenamefont {Yu}, \citenamefont {Bradley}, \citenamefont {Baker}, \citenamefont {Bell}, \citenamefont {Rubinsztein-Dunlop}, \citenamefont {Davis},\ and\ \citenamefont {Neely}}]{gauthier2019giant}%
  \BibitemOpen
  \bibfield  {author} {\bibinfo {author} {\bibfnamefont {G.}~\bibnamefont {Gauthier}}, \bibinfo {author} {\bibfnamefont {M.~T.}\ \bibnamefont {Reeves}}, \bibinfo {author} {\bibfnamefont {X.}~\bibnamefont {Yu}}, \bibinfo {author} {\bibfnamefont {A.~S.}\ \bibnamefont {Bradley}}, \bibinfo {author} {\bibfnamefont {M.~A.}\ \bibnamefont {Baker}}, \bibinfo {author} {\bibfnamefont {T.~A.}\ \bibnamefont {Bell}}, \bibinfo {author} {\bibfnamefont {H.}~\bibnamefont {Rubinsztein-Dunlop}}, \bibinfo {author} {\bibfnamefont {M.~J.}\ \bibnamefont {Davis}},\ and\ \bibinfo {author} {\bibfnamefont {T.~W.}\ \bibnamefont {Neely}},\ }\bibfield  {title} {\bibinfo {title} {Giant vortex clusters in a two-dimensional quantum fluid},\ }\href@noop {} {\bibfield  {journal} {\bibinfo  {journal} {Science}\ }\textbf {\bibinfo {volume} {364}},\ \bibinfo {pages} {1264} (\bibinfo {year} {2019})}\BibitemShut {NoStop}%
\bibitem [{\citenamefont {Johnstone}\ \emph {et~al.}(2019)\citenamefont {Johnstone}, \citenamefont {Groszek}, \citenamefont {Starkey}, \citenamefont {Billington}, \citenamefont {Simula},\ and\ \citenamefont {Helmerson}}]{johnstone2019evolution}%
  \BibitemOpen
  \bibfield  {author} {\bibinfo {author} {\bibfnamefont {S.~P.}\ \bibnamefont {Johnstone}}, \bibinfo {author} {\bibfnamefont {A.~J.}\ \bibnamefont {Groszek}}, \bibinfo {author} {\bibfnamefont {P.~T.}\ \bibnamefont {Starkey}}, \bibinfo {author} {\bibfnamefont {C.~J.}\ \bibnamefont {Billington}}, \bibinfo {author} {\bibfnamefont {T.~P.}\ \bibnamefont {Simula}},\ and\ \bibinfo {author} {\bibfnamefont {K.}~\bibnamefont {Helmerson}},\ }\bibfield  {title} {\bibinfo {title} {Evolution of large-scale flow from turbulence in a two-dimensional superfluid},\ }\href@noop {} {\bibfield  {journal} {\bibinfo  {journal} {Science}\ }\textbf {\bibinfo {volume} {364}},\ \bibinfo {pages} {1267} (\bibinfo {year} {2019})}\BibitemShut {NoStop}%
\bibitem [{\citenamefont {Reeves}\ \emph {et~al.}(2013)\citenamefont {Reeves}, \citenamefont {Billam}, \citenamefont {Anderson},\ and\ \citenamefont {Bradley}}]{reeves2013inverse}%
  \BibitemOpen
  \bibfield  {author} {\bibinfo {author} {\bibfnamefont {M.~T.}\ \bibnamefont {Reeves}}, \bibinfo {author} {\bibfnamefont {T.~P.}\ \bibnamefont {Billam}}, \bibinfo {author} {\bibfnamefont {B.~P.}\ \bibnamefont {Anderson}},\ and\ \bibinfo {author} {\bibfnamefont {A.~S.}\ \bibnamefont {Bradley}},\ }\bibfield  {title} {\bibinfo {title} {Inverse energy cascade in forced two-dimensional quantum turbulence},\ }\href@noop {} {\bibfield  {journal} {\bibinfo  {journal} {Physical review letters}\ }\textbf {\bibinfo {volume} {110}},\ \bibinfo {pages} {104501} (\bibinfo {year} {2013})}\BibitemShut {NoStop}%
\bibitem [{\citenamefont {del Valle-Inclan~Redondo}\ \emph {et~al.}(2023)\citenamefont {del Valle-Inclan~Redondo}, \citenamefont {Schneider}, \citenamefont {Klembt}, \citenamefont {Hoofling}, \citenamefont {Tarucha},\ and\ \citenamefont {Fraser}}]{del2023optically}%
  \BibitemOpen
  \bibfield  {author} {\bibinfo {author} {\bibfnamefont {Y.}~\bibnamefont {del Valle-Inclan~Redondo}}, \bibinfo {author} {\bibfnamefont {C.}~\bibnamefont {Schneider}}, \bibinfo {author} {\bibfnamefont {S.}~\bibnamefont {Klembt}}, \bibinfo {author} {\bibfnamefont {S.}~\bibnamefont {Hoofling}}, \bibinfo {author} {\bibfnamefont {S.}~\bibnamefont {Tarucha}},\ and\ \bibinfo {author} {\bibfnamefont {M.~D.}\ \bibnamefont {Fraser}},\ }\bibfield  {title} {\bibinfo {title} {Optically driven rotation of exciton--polariton condensates},\ }\href@noop {} {\bibfield  {journal} {\bibinfo  {journal} {Nano Letters}\ }\textbf {\bibinfo {volume} {23}},\ \bibinfo {pages} {4564} (\bibinfo {year} {2023})}\BibitemShut {NoStop}%
\bibitem [{\citenamefont {Gnusov}\ \emph {et~al.}(2023)\citenamefont {Gnusov}, \citenamefont {Harrison}, \citenamefont {Alyatkin}, \citenamefont {Sitnik}, \citenamefont {T{\"o}pfer}, \citenamefont {Sigurdsson},\ and\ \citenamefont {Lagoudakis}}]{gnusov2023quantum}%
  \BibitemOpen
  \bibfield  {author} {\bibinfo {author} {\bibfnamefont {I.}~\bibnamefont {Gnusov}}, \bibinfo {author} {\bibfnamefont {S.}~\bibnamefont {Harrison}}, \bibinfo {author} {\bibfnamefont {S.}~\bibnamefont {Alyatkin}}, \bibinfo {author} {\bibfnamefont {K.}~\bibnamefont {Sitnik}}, \bibinfo {author} {\bibfnamefont {J.}~\bibnamefont {T{\"o}pfer}}, \bibinfo {author} {\bibfnamefont {H.}~\bibnamefont {Sigurdsson}},\ and\ \bibinfo {author} {\bibfnamefont {P.}~\bibnamefont {Lagoudakis}},\ }\bibfield  {title} {\bibinfo {title} {Quantum vortex formation in the ``rotating bucket'' experiment with polariton condensates},\ }\href@noop {} {\bibfield  {journal} {\bibinfo  {journal} {Science Advances}\ }\textbf {\bibinfo {volume} {9}},\ \bibinfo {pages} {eadd1299} (\bibinfo {year} {2023})}\BibitemShut {NoStop}%
\bibitem [{\citenamefont {Gnusov}\ \emph {et~al.}(2024)\citenamefont {Gnusov}, \citenamefont {Harrison}, \citenamefont {Alyatkin}, \citenamefont {Sitnik}, \citenamefont {Sigurdsson},\ and\ \citenamefont {Lagoudakis}}]{gnusov2024vortex}%
  \BibitemOpen
  \bibfield  {author} {\bibinfo {author} {\bibfnamefont {I.}~\bibnamefont {Gnusov}}, \bibinfo {author} {\bibfnamefont {S.}~\bibnamefont {Harrison}}, \bibinfo {author} {\bibfnamefont {S.}~\bibnamefont {Alyatkin}}, \bibinfo {author} {\bibfnamefont {K.}~\bibnamefont {Sitnik}}, \bibinfo {author} {\bibfnamefont {H.}~\bibnamefont {Sigurdsson}},\ and\ \bibinfo {author} {\bibfnamefont {P.~G.}\ \bibnamefont {Lagoudakis}},\ }\bibfield  {title} {\bibinfo {title} {Vortex clusters in a stirred polariton condensate},\ }\href@noop {} {\bibfield  {journal} {\bibinfo  {journal} {Physical Review B}\ }\textbf {\bibinfo {volume} {109}},\ \bibinfo {pages} {104503} (\bibinfo {year} {2024})}\BibitemShut {NoStop}%
\bibitem [{\citenamefont {De~Giorgi}\ \emph {et~al.}(2014)\citenamefont {De~Giorgi}, \citenamefont {Ballarini}, \citenamefont {Cazzato}, \citenamefont {Deligeorgis}, \citenamefont {Tsintzos}, \citenamefont {Hatzopoulos}, \citenamefont {Savvidis}, \citenamefont {Gigli}, \citenamefont {Laussy},\ and\ \citenamefont {Sanvitto}}]{de2014relaxation}%
  \BibitemOpen
  \bibfield  {author} {\bibinfo {author} {\bibfnamefont {M.}~\bibnamefont {De~Giorgi}}, \bibinfo {author} {\bibfnamefont {D.}~\bibnamefont {Ballarini}}, \bibinfo {author} {\bibfnamefont {P.}~\bibnamefont {Cazzato}}, \bibinfo {author} {\bibfnamefont {G.}~\bibnamefont {Deligeorgis}}, \bibinfo {author} {\bibfnamefont {S.~I.}\ \bibnamefont {Tsintzos}}, \bibinfo {author} {\bibfnamefont {Z.}~\bibnamefont {Hatzopoulos}}, \bibinfo {author} {\bibfnamefont {P.~G.}\ \bibnamefont {Savvidis}}, \bibinfo {author} {\bibfnamefont {G.}~\bibnamefont {Gigli}}, \bibinfo {author} {\bibfnamefont {F.~P.}\ \bibnamefont {Laussy}},\ and\ \bibinfo {author} {\bibfnamefont {D.}~\bibnamefont {Sanvitto}},\ }\bibfield  {title} {\bibinfo {title} {Relaxation oscillations in the formation of a polariton condensate},\ }\href@noop {} {\bibfield  {journal} {\bibinfo  {journal} {Physical Review Letters}\ }\textbf {\bibinfo {volume} {112}},\ \bibinfo {pages} {113602} (\bibinfo {year} {2014})}\BibitemShut {NoStop}%
\bibitem [{\citenamefont {Park}\ \emph {et~al.}(2025)\citenamefont {Park}, \citenamefont {Koniakhin}, \citenamefont {Choi}, \citenamefont {Choi}, \citenamefont {Park}, \citenamefont {Park}, \citenamefont {Song}, \citenamefont {Cho},\ and\ \citenamefont {Choi}}]{park2025exciton}%
  \BibitemOpen
  \bibfield  {author} {\bibinfo {author} {\bibfnamefont {M.}~\bibnamefont {Park}}, \bibinfo {author} {\bibfnamefont {S.}~\bibnamefont {Koniakhin}}, \bibinfo {author} {\bibfnamefont {S.}~\bibnamefont {Choi}}, \bibinfo {author} {\bibfnamefont {D.}~\bibnamefont {Choi}}, \bibinfo {author} {\bibfnamefont {S.~I.}\ \bibnamefont {Park}}, \bibinfo {author} {\bibfnamefont {S.}~\bibnamefont {Park}}, \bibinfo {author} {\bibfnamefont {J.~D.}\ \bibnamefont {Song}}, \bibinfo {author} {\bibfnamefont {Y.-H.}\ \bibnamefont {Cho}},\ and\ \bibinfo {author} {\bibfnamefont {H.}~\bibnamefont {Choi}},\ }\bibfield  {title} {\bibinfo {title} {Exciton reservoir-induced destabilization and reformation of polariton condensate},\ }\href@noop {} {\bibfield  {journal} {\bibinfo  {journal} {Optics Express}\ }\textbf {\bibinfo {volume} {33}},\ \bibinfo {pages} {18530} (\bibinfo {year} {2025})}\BibitemShut {NoStop}%
\bibitem [{\citenamefont {Boulier}\ \emph {et~al.}(2015)\citenamefont {Boulier}, \citenamefont {Ter{\c{c}}as}, \citenamefont {Solnyshkov}, \citenamefont {Glorieux}, \citenamefont {Giacobino}, \citenamefont {Malpuech},\ and\ \citenamefont {Bramati}}]{boulier2015vortex}%
  \BibitemOpen
  \bibfield  {author} {\bibinfo {author} {\bibfnamefont {T.}~\bibnamefont {Boulier}}, \bibinfo {author} {\bibfnamefont {H.}~\bibnamefont {Ter{\c{c}}as}}, \bibinfo {author} {\bibfnamefont {D.}~\bibnamefont {Solnyshkov}}, \bibinfo {author} {\bibfnamefont {Q.}~\bibnamefont {Glorieux}}, \bibinfo {author} {\bibfnamefont {E.}~\bibnamefont {Giacobino}}, \bibinfo {author} {\bibfnamefont {G.}~\bibnamefont {Malpuech}},\ and\ \bibinfo {author} {\bibfnamefont {A.}~\bibnamefont {Bramati}},\ }\bibfield  {title} {\bibinfo {title} {Vortex chain in a resonantly pumped polariton superfluid},\ }\href@noop {} {\bibfield  {journal} {\bibinfo  {journal} {Scientific reports}\ }\textbf {\bibinfo {volume} {5}},\ \bibinfo {pages} {9230} (\bibinfo {year} {2015})}\BibitemShut {NoStop}%
\bibitem [{\citenamefont {Panico}\ \emph {et~al.}(2021)\citenamefont {Panico}, \citenamefont {Macorini}, \citenamefont {Dominici}, \citenamefont {Gianfrate}, \citenamefont {Fieramosca}, \citenamefont {De~Giorgi}, \citenamefont {Gigli}, \citenamefont {Sanvitto}, \citenamefont {Lanotte},\ and\ \citenamefont {Ballarini}}]{panico2021dynamics}%
  \BibitemOpen
  \bibfield  {author} {\bibinfo {author} {\bibfnamefont {R.}~\bibnamefont {Panico}}, \bibinfo {author} {\bibfnamefont {G.}~\bibnamefont {Macorini}}, \bibinfo {author} {\bibfnamefont {L.}~\bibnamefont {Dominici}}, \bibinfo {author} {\bibfnamefont {A.}~\bibnamefont {Gianfrate}}, \bibinfo {author} {\bibfnamefont {A.}~\bibnamefont {Fieramosca}}, \bibinfo {author} {\bibfnamefont {M.}~\bibnamefont {De~Giorgi}}, \bibinfo {author} {\bibfnamefont {G.}~\bibnamefont {Gigli}}, \bibinfo {author} {\bibfnamefont {D.}~\bibnamefont {Sanvitto}}, \bibinfo {author} {\bibfnamefont {A.~S.}\ \bibnamefont {Lanotte}},\ and\ \bibinfo {author} {\bibfnamefont {D.}~\bibnamefont {Ballarini}},\ }\bibfield  {title} {\bibinfo {title} {Dynamics of a vortex lattice in an expanding polariton quantum fluid},\ }\href@noop {} {\bibfield  {journal} {\bibinfo  {journal} {Physical Review Letters}\ }\textbf {\bibinfo {volume} {127}},\ \bibinfo {pages} {047401} (\bibinfo {year} {2021})}\BibitemShut {NoStop}%
\bibitem [{\citenamefont {Burger}\ \emph {et~al.}(1999)\citenamefont {Burger}, \citenamefont {Bongs}, \citenamefont {Dettmer}, \citenamefont {Ertmer}, \citenamefont {Sengstock}, \citenamefont {Sanpera}, \citenamefont {Shlyapnikov},\ and\ \citenamefont {Lewenstein}}]{burger1999dark}%
  \BibitemOpen
  \bibfield  {author} {\bibinfo {author} {\bibfnamefont {S.}~\bibnamefont {Burger}}, \bibinfo {author} {\bibfnamefont {K.}~\bibnamefont {Bongs}}, \bibinfo {author} {\bibfnamefont {S.}~\bibnamefont {Dettmer}}, \bibinfo {author} {\bibfnamefont {W.}~\bibnamefont {Ertmer}}, \bibinfo {author} {\bibfnamefont {K.}~\bibnamefont {Sengstock}}, \bibinfo {author} {\bibfnamefont {A.}~\bibnamefont {Sanpera}}, \bibinfo {author} {\bibfnamefont {G.~V.}\ \bibnamefont {Shlyapnikov}},\ and\ \bibinfo {author} {\bibfnamefont {M.}~\bibnamefont {Lewenstein}},\ }\bibfield  {title} {\bibinfo {title} {Dark solitons in bose-einstein condensates},\ }\href@noop {} {\bibfield  {journal} {\bibinfo  {journal} {Physical Review Letters}\ }\textbf {\bibinfo {volume} {83}},\ \bibinfo {pages} {5198} (\bibinfo {year} {1999})}\BibitemShut {NoStop}%
\bibitem [{\citenamefont {Comaron}\ \emph {et~al.}(2025)\citenamefont {Comaron}, \citenamefont {Panico}, \citenamefont {Ballarini},\ and\ \citenamefont {Matuszewski}}]{comaron2024dynamics}%
  \BibitemOpen
  \bibfield  {author} {\bibinfo {author} {\bibfnamefont {P.}~\bibnamefont {Comaron}}, \bibinfo {author} {\bibfnamefont {R.}~\bibnamefont {Panico}}, \bibinfo {author} {\bibfnamefont {D.}~\bibnamefont {Ballarini}},\ and\ \bibinfo {author} {\bibfnamefont {M.}~\bibnamefont {Matuszewski}},\ }\bibfield  {title} {\bibinfo {title} {Dynamics of onsager vortex clustering in decaying turbulent polariton quantum fluids},\ }\href@noop {} {\bibfield  {journal} {\bibinfo  {journal} {Physical Review Research}\ }\textbf {\bibinfo {volume} {7}},\ \bibinfo {pages} {L022006} (\bibinfo {year} {2025})}\BibitemShut {NoStop}%
\bibitem [{\citenamefont {Solnyshkov}\ \emph {et~al.}(2014)\citenamefont {Solnyshkov}, \citenamefont {Tercas}, \citenamefont {Dini},\ and\ \citenamefont {Malpuech}}]{Solnyshkov2014}%
  \BibitemOpen
  \bibfield  {author} {\bibinfo {author} {\bibfnamefont {D.~D.}\ \bibnamefont {Solnyshkov}}, \bibinfo {author} {\bibfnamefont {H.}~\bibnamefont {Tercas}}, \bibinfo {author} {\bibfnamefont {K.}~\bibnamefont {Dini}},\ and\ \bibinfo {author} {\bibfnamefont {G.}~\bibnamefont {Malpuech}},\ }\bibfield  {title} {\bibinfo {title} {Hybrid {B}oltzmann–{G}ross-{P}itaevskii theory of {B}ose-einstein condensation and superfluidity in open driven-dissipative systems},\ }\href {https://doi.org/https://doi.org/10.1103/PhysRevA.89.033626} {\bibfield  {journal} {\bibinfo  {journal} {Physical Review A}\ }\textbf {\bibinfo {volume} {89}},\ \bibinfo {pages} {033626} (\bibinfo {year} {2014})}\BibitemShut {NoStop}%
\bibitem [{\citenamefont {Boulier}\ \emph {et~al.}(2016)\citenamefont {Boulier}, \citenamefont {Cancellieri}, \citenamefont {Sangouard}, \citenamefont {Glorieux}, \citenamefont {Kavokin}, \citenamefont {Whittaker}, \citenamefont {Giacobino},\ and\ \citenamefont {Bramati}}]{boulier2016injection}%
  \BibitemOpen
  \bibfield  {author} {\bibinfo {author} {\bibfnamefont {T.}~\bibnamefont {Boulier}}, \bibinfo {author} {\bibfnamefont {E.}~\bibnamefont {Cancellieri}}, \bibinfo {author} {\bibfnamefont {N.~D.}\ \bibnamefont {Sangouard}}, \bibinfo {author} {\bibfnamefont {Q.}~\bibnamefont {Glorieux}}, \bibinfo {author} {\bibfnamefont {A.}~\bibnamefont {Kavokin}}, \bibinfo {author} {\bibfnamefont {D.~M.}\ \bibnamefont {Whittaker}}, \bibinfo {author} {\bibfnamefont {E.}~\bibnamefont {Giacobino}},\ and\ \bibinfo {author} {\bibfnamefont {A.}~\bibnamefont {Bramati}},\ }\bibfield  {title} {\bibinfo {title} {Injection of orbital angular momentum and storage of quantized vortices in polariton superfluids},\ }\href@noop {} {\bibfield  {journal} {\bibinfo  {journal} {Physical review letters}\ }\textbf {\bibinfo {volume} {116}},\ \bibinfo {pages} {116402} (\bibinfo {year} {2016})}\BibitemShut {NoStop}%
\bibitem [{\citenamefont {Koniakhin}\ \emph {et~al.}(2020)\citenamefont {Koniakhin}, \citenamefont {Bleu}, \citenamefont {Malpuech},\ and\ \citenamefont {Solnyshkov}}]{Koniakhin2020}%
  \BibitemOpen
  \bibfield  {author} {\bibinfo {author} {\bibfnamefont {S.}~\bibnamefont {Koniakhin}}, \bibinfo {author} {\bibfnamefont {O.}~\bibnamefont {Bleu}}, \bibinfo {author} {\bibfnamefont {G.}~\bibnamefont {Malpuech}},\ and\ \bibinfo {author} {\bibfnamefont {D.}~\bibnamefont {Solnyshkov}},\ }\bibfield  {title} {\bibinfo {title} {2{D} quantum turbulence in a polariton quantum fluid},\ }\href {https://doi.org/https://doi.org/10.1016/j.chaos.2019.109574} {\bibfield  {journal} {\bibinfo  {journal} {Chaos, Solitons \& Fractals}\ }\textbf {\bibinfo {volume} {132}},\ \bibinfo {pages} {109574} (\bibinfo {year} {2020})}\BibitemShut {NoStop}%
\bibitem [{\citenamefont {Kraichnan}(1967)}]{kraichnan1967inertial}%
  \BibitemOpen
  \bibfield  {author} {\bibinfo {author} {\bibfnamefont {R.~H.}\ \bibnamefont {Kraichnan}},\ }\bibfield  {title} {\bibinfo {title} {Inertial ranges in two-dimensional turbulence},\ }\href@noop {} {\bibfield  {journal} {\bibinfo  {journal} {Physics of fluids}\ }\textbf {\bibinfo {volume} {10}},\ \bibinfo {pages} {1417} (\bibinfo {year} {1967})}\BibitemShut {NoStop}%
\bibitem [{\citenamefont {Groszek}\ \emph {et~al.}(2018)\citenamefont {Groszek}, \citenamefont {Davis}, \citenamefont {Paganin}, \citenamefont {Helmerson},\ and\ \citenamefont {Simula}}]{groszek2018vortex}%
  \BibitemOpen
  \bibfield  {author} {\bibinfo {author} {\bibfnamefont {A.~J.}\ \bibnamefont {Groszek}}, \bibinfo {author} {\bibfnamefont {M.~J.}\ \bibnamefont {Davis}}, \bibinfo {author} {\bibfnamefont {D.~M.}\ \bibnamefont {Paganin}}, \bibinfo {author} {\bibfnamefont {K.}~\bibnamefont {Helmerson}},\ and\ \bibinfo {author} {\bibfnamefont {T.~P.}\ \bibnamefont {Simula}},\ }\bibfield  {title} {\bibinfo {title} {Vortex thermometry for turbulent two-dimensional fluids},\ }\href@noop {} {\bibfield  {journal} {\bibinfo  {journal} {Physical review letters}\ }\textbf {\bibinfo {volume} {120}},\ \bibinfo {pages} {034504} (\bibinfo {year} {2018})}\BibitemShut {NoStop}%
\bibitem [{\citenamefont {M{\"u}ller}\ and\ \citenamefont {Krstulovic}(2024)}]{muller2024exploring}%
  \BibitemOpen
  \bibfield  {author} {\bibinfo {author} {\bibfnamefont {N.~P.}\ \bibnamefont {M{\"u}ller}}\ and\ \bibinfo {author} {\bibfnamefont {G.}~\bibnamefont {Krstulovic}},\ }\bibfield  {title} {\bibinfo {title} {Exploring the equivalence between two-dimensional classical and quantum turbulence through velocity circulation statistics},\ }\href@noop {} {\bibfield  {journal} {\bibinfo  {journal} {Physical Review Letters}\ }\textbf {\bibinfo {volume} {132}},\ \bibinfo {pages} {094002} (\bibinfo {year} {2024})}\BibitemShut {NoStop}%
\bibitem [{\citenamefont {Berman}\ \emph {et~al.}(2015)\citenamefont {Berman}, \citenamefont {Kezerashvili}, \citenamefont {Kolmakov},\ and\ \citenamefont {Pomirchi}}]{berman2015spontaneous}%
  \BibitemOpen
  \bibfield  {author} {\bibinfo {author} {\bibfnamefont {O.~L.}\ \bibnamefont {Berman}}, \bibinfo {author} {\bibfnamefont {R.~Y.}\ \bibnamefont {Kezerashvili}}, \bibinfo {author} {\bibfnamefont {G.~V.}\ \bibnamefont {Kolmakov}},\ and\ \bibinfo {author} {\bibfnamefont {L.~M.}\ \bibnamefont {Pomirchi}},\ }\bibfield  {title} {\bibinfo {title} {Spontaneous formation and nonequilibrium dynamics of a soliton-shaped bose-einstein condensate in a trap},\ }\href@noop {} {\bibfield  {journal} {\bibinfo  {journal} {Physical Review E}\ }\textbf {\bibinfo {volume} {91}},\ \bibinfo {pages} {062901} (\bibinfo {year} {2015})}\BibitemShut {NoStop}%
\bibitem [{\citenamefont {Berman}\ \emph {et~al.}(2012)\citenamefont {Berman}, \citenamefont {Kezerashvili}, \citenamefont {Kolmakov},\ and\ \citenamefont {Lozovik}}]{berman2012turbulence}%
  \BibitemOpen
  \bibfield  {author} {\bibinfo {author} {\bibfnamefont {O.}~\bibnamefont {Berman}}, \bibinfo {author} {\bibfnamefont {R.~Y.}\ \bibnamefont {Kezerashvili}}, \bibinfo {author} {\bibfnamefont {G.}~\bibnamefont {Kolmakov}},\ and\ \bibinfo {author} {\bibfnamefont {Y.~E.}\ \bibnamefont {Lozovik}},\ }\bibfield  {title} {\bibinfo {title} {Turbulence in a bose-einstein condensate of dipolar excitons in coupled quantum wells},\ }\href@noop {} {\bibfield  {journal} {\bibinfo  {journal} {Physical Review B—Condensed Matter and Materials Physics}\ }\textbf {\bibinfo {volume} {86}},\ \bibinfo {pages} {045108} (\bibinfo {year} {2012})}\BibitemShut {NoStop}%
\bibitem [{\citenamefont {Nelsen}\ \emph {et~al.}(2013)\citenamefont {Nelsen}, \citenamefont {Liu}, \citenamefont {Steger}, \citenamefont {Snoke}, \citenamefont {Balili}, \citenamefont {West},\ and\ \citenamefont {Pfeiffer}}]{nelsen2013dissipationless}%
  \BibitemOpen
  \bibfield  {author} {\bibinfo {author} {\bibfnamefont {B.}~\bibnamefont {Nelsen}}, \bibinfo {author} {\bibfnamefont {G.}~\bibnamefont {Liu}}, \bibinfo {author} {\bibfnamefont {M.}~\bibnamefont {Steger}}, \bibinfo {author} {\bibfnamefont {D.~W.}\ \bibnamefont {Snoke}}, \bibinfo {author} {\bibfnamefont {R.}~\bibnamefont {Balili}}, \bibinfo {author} {\bibfnamefont {K.}~\bibnamefont {West}},\ and\ \bibinfo {author} {\bibfnamefont {L.}~\bibnamefont {Pfeiffer}},\ }\bibfield  {title} {\bibinfo {title} {Dissipationless flow and sharp threshold of a polariton condensate with long lifetime},\ }\href@noop {} {\bibfield  {journal} {\bibinfo  {journal} {Physical Review X}\ }\textbf {\bibinfo {volume} {3}},\ \bibinfo {pages} {041015} (\bibinfo {year} {2013})}\BibitemShut {NoStop}%
\end{thebibliography}%

\newpage
\appendix
\onecolumngrid



\section{NUMERICAL SIMULATIONS}

The investigation the efficiency of the wave function imprinting schemes as turbulence excitation techniques in comparison with spoon-stirring procedure is based on numerical simulation of the following driven-dissipative Gross-Pitaevskii equation:
\begin{equation}
    \label{eq: modified GPE}
    i\hbar\frac{\partial}{\partial t} \psi(\mathbf{r}, t) = \left[\left(i\beta(t)-1\right)\frac{\hbar^2\nabla^2}{2m}+g|\psi(\mathbf{r},t)|^2+ V(\mathbf{r},t)-i\frac{\hbar}{2}\gamma(t)\right]\psi(\mathbf{r},t).
\end{equation}

The considered polariton system is characterized as follows: polariton mass $m=5\times10^{-5} \ m_0$ (bare electron mass $m_0= 9.1\times10^{-28}$~g), interaction constant $g = 5\times10^{-3}~\mathrm{meV}\mu \mathrm{m}^2$. The time integration is solved by means of third order Adam-Bashforth method, with time step $\Delta t=0.004$~ps, total duration $T_{tot}=4.56$~ns and a square mesh grid of size $L=512 \ \mu$m. For the calculation of the Laplace operator (kinetic energy term), we used the Fourier transform with massive parallelization of matrix operations provided by GPU in \textsc{PyTorch}. 

Within this most general formulation, the dynamical equation above enables to perform analysis for both ``standard'' stirring techniques (widely used in atomic Bose-Einstein condensates) based for a time-dependent rotating potential and phase-pattern imprinting schemes with stationary potential, $V(\mathbf{r},t)= V(\mathbf{r})$. The confining circular hard wall potential of 480~$\mu$m in diameter, $V(\mathbf{r})$, was normally used. Moreover, all the considered simulations consist of an initial preparatory stage, especially required by the spoon stirring method, and an equally-long analysis stage to investigate quantum turbulence.

The simulation of spoon-stirring was carried out by a 10~$\mu$m propagating potential with a diameter $d=12 \ \mu m$ and rotating at 1.3~$\mu$m/ps with an orbit radius $R_{sp}= 160 \ \mu m$ during the preparatory phase (3 complete spoon rotations).
Considering a depth $V_{max}=10\ meV$ and rotational period $T_0 =760$~ps, the total potential  $V(\mathbf{r},t)=V(\mathbf{r})+V_{st}(\mathbf{r},t)$ is given by a stirring potential defined as  
$V_{st}(\mathbf{r},t) = A(t)\cdot H(D/2-|\mathbf{r}-\mathbf{r}_{sp}(t)|)$, with $H$ Heaviside step function and 
\begin{equation}
\label{eq: spoon potential}
    \begin{split}
    A(t)=V_{max}\cdot\begin{cases}
        t/T_0 \  & t\leq T_0= T_{tot}/6,
        \\ 
        1\  & T_0< t\leq 2T_0,
        \\
        (3T_0-t)/T_0 \  & 2T_0< t\leq 3T_0,
        \\
        0 \ &  t>3T_0= T_{tot}/2 
    \end{cases}
    \ \ ,\ \ \mathbf{r}_{sp}(t)=L/2+R_{sp}\left[ \cos(2\pi t/T_0),\ \sin(2\pi t/T_0) \right].
    \end{split}
\end{equation}

The time-modulation amplitude function $A(t)$ is considered to smoothly stir on time the condensate during the whole stirring interval, avoiding to generate excessive density noise. For such a case the initial configuration is $\psi_0$ is an homogeneous state with density $n_0 =|\psi_0|^2$. For averaging, the spoon-stirring configurations, we have changed the orbit radius (within 1\%).

As for the imprinting methods, two different families of phase pattern were considered: Tile and Pixel Imprinting schemes. In the first case, we divide the wave function into the 8x8 grid of the square tiles of size 64~$\mu$m and fill each of them with a randomly oriented plane wave with wave length 12~$\mu$m (Tile-12), 8~$\mu$m (Tile-8) or 4~$\mu$m (Tile-4). For the sake of completeness, this imprint method was also studied also 32x32, 16x16 and 4x4 grids with square size $l_b$ respectively equal to 16~$\mu$m, 32~$\mu$m and 128~$\mu$m. 
As the condensate evolves in time, vortices are randomly creates at the borders of the tiles by the means of only wave function phase pattern (no time-dependent rotating potential).
With a similar principle, in the second case we divide the wave function into pixels of the size 8~$\mu$m (Pixel-8) or 4~$\mu$m (Pixel-4) of the random phase each.


\section{\label{sec: links original sims}
links to animation}
Here in the present section, we provide a list of links to movies of time evolution of the wave function $\psi(\mathbf{r},t)$, by means of the density field $|\psi(\mathbf{r},t)|^2$ and phase field of $\psi(\mathbf{r},t)\ |\psi(\mathbf{r_c},t)|/\psi(\mathbf{r_c},t)$ (phase rescaled such that it is zero in the center $\mathbf{r_c}$). The links accessed August 10th, 2025.

\subsection{Conservative case and finite lifetime}

Conservative case of long-living polaritons with  \underline{$\tau\gg1$~ns}:
\begin{itemize}
    \item \href{https://youtu.be/ncXklJiJrv0}{Spoon-Stirring}
    \item \href{https://youtu.be/DXI57H1XQMY}{Tile-4 imprinting scheme}
    \item \href{https://youtu.be/5ONPTJnGQOo}{Tile-8 imprinting scheme}
    \item \href{https://youtu.be/BUS5jgQK-Ro}{Tile-12 imprinting scheme}
    \item \href{https://youtu.be/a_xgVLHlfUo}{Pixel-4 imprinting scheme}
    \item \href{https://youtu.be/I4UVjhaLQoc}{Pixel-8 imprinting scheme}
\end{itemize}

Decaying polariton case with finite lifetime \underline{$\tau=1.89$~ns}:
\begin{itemize}
    \item \href{https://youtu.be/pA0cKXpR8KA}{Spoon-Stirring}
    \item \href{https://youtu.be/n93jJ2VHZEE}{Tile-4 imprinting scheme}
    \item \href{https://youtu.be/rkPQvFIGei0}{Tile-8 imprinting scheme}
    \item \href{https://youtu.be/BUS5jgQK-Ro}{Tile-12 imprinting scheme}
    \item \href{https://youtu.be/A4ofMPlfado}{Pixel-4 imprinting scheme}
    \item \href{https://youtu.be/wKPCjAbyYSw}{Pixel-8 imprinting scheme}
\end{itemize}

\subsection{Simulations without high-energy wave damping ($\beta = 0$)}
\begin{itemize}
    \item \href{https://youtu.be/g_uJ3fDqCWA}{Tile-8 imprinting Scheme, $\tau \gg1$~ns}
    \item \href{https://youtu.be/hTMF-1H5mgg}{Tile-8 imprinting Scheme, $\tau =1.89$~ns}
\end{itemize}

\subsection{Simulations Tile-8 imprinting scheme for different tile grids}
\begin{itemize}
    \item \href{https://youtu.be/aomYBw7BBvU}{$4\times4$ Grid, block size $l_b = 128~\mu$m}
    \item \href{https://youtu.be/WUWNu4mPuso}{$8\times8$ Grid, block size $l_b = 64~\mu$m}
    \item \href{https://youtu.be/4nnERQS7uQ8}{$16\times16$ Grid, block size $l_b = 32~\mu$m}
    \item \href{https://youtu.be/GrFq7nyLrQ4}{$32\times32$ Grid, block size $l_b = 16~\mu$m}
\end{itemize}


\newpage

\section{Simulation snapshots for spoon stirring and Tile-8 at various time moments}

\begin{figure}[h!]
    \begin{center}
        \includegraphics[width=0.9\linewidth]{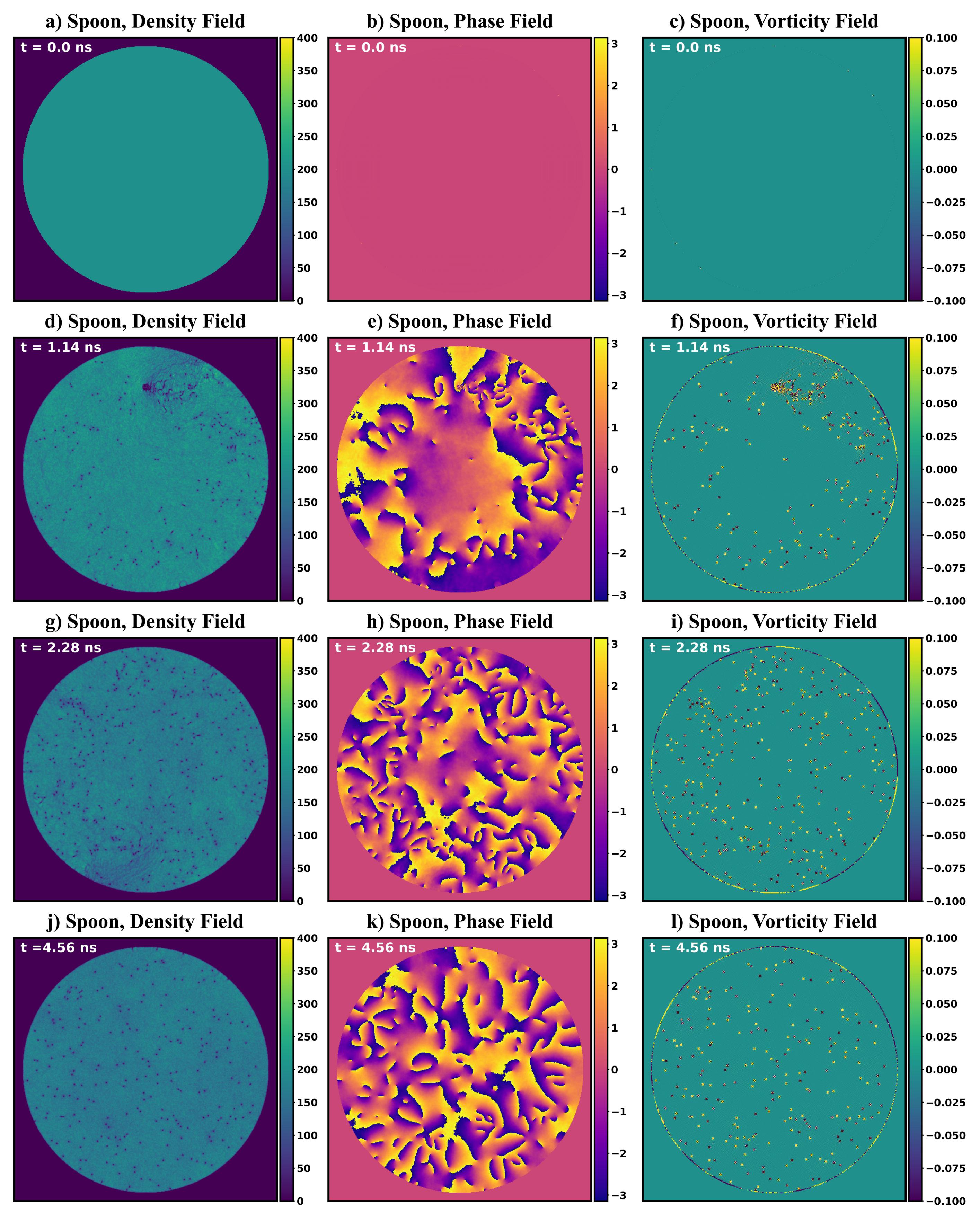}
    \end{center}
    
    \caption{\label{fig_suppl_spoon} Simulation snapshots for \underline{spoon stirring} for different time moments, illustrating the initial stirring process with $\beta=0.001$, panels \textbf{a-f)}, and the condensate free time evolution (analysis stage with $\beta=0$), panels \textbf{g-l)}. The simulation considers long-living polaritons (conservative case), and the full video is available online from section \ref{sec: links original sims} 
    }
\end{figure}

\begin{figure}[h!]
    \begin{center}
        \includegraphics[width=0.9\linewidth]{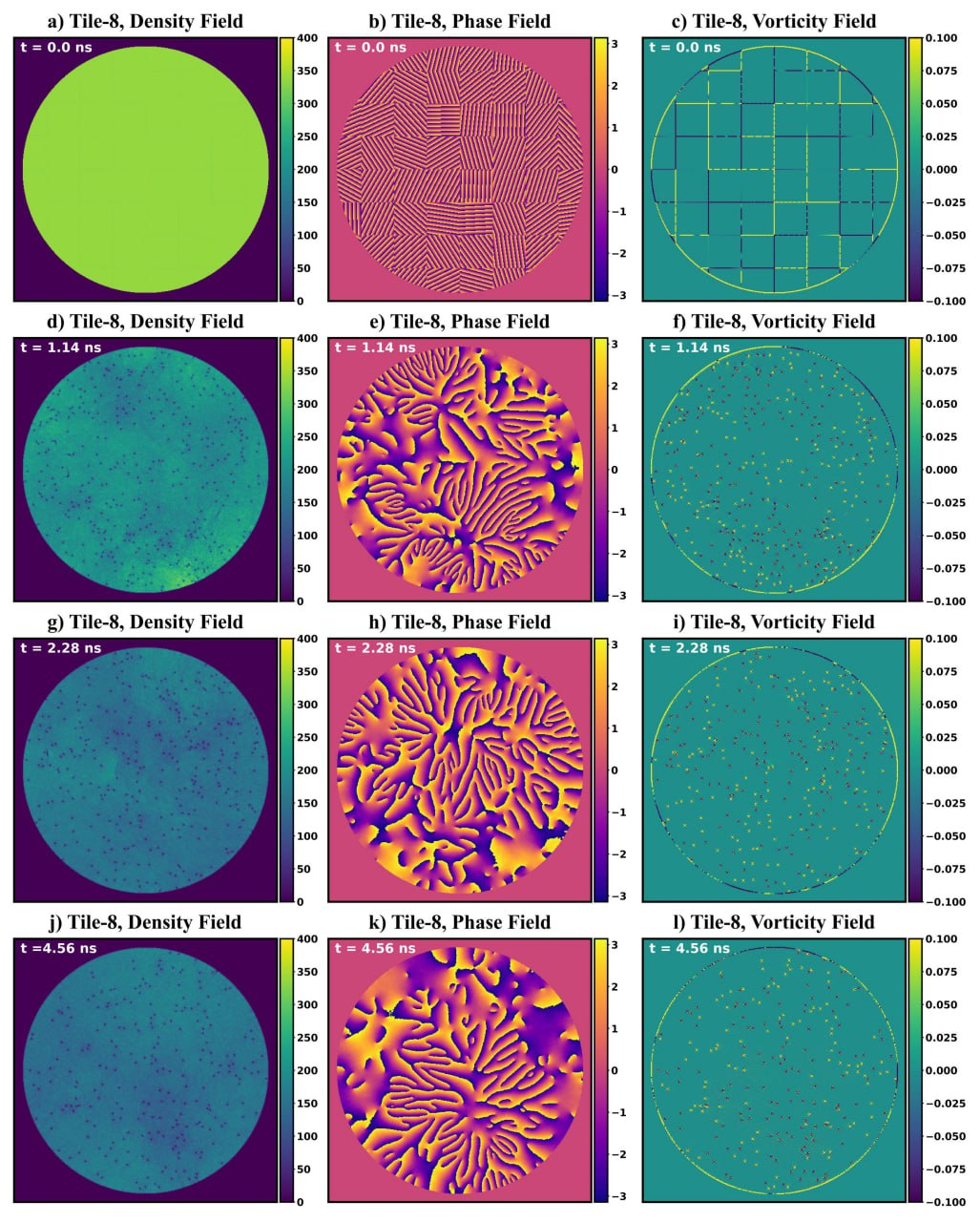}
    \end{center}
    \caption{\label{fig_suppl_tile-8}Simulation snapshots for \underline{Tile-8 imprinting} scheme for different time moments for an initial $8\times8$ tile grid (block size $l_b = 64~\mu$m) and imprinted plane wave of 8~$\mu$m wave length. Panels \textbf{a-c)} illustrate the fields of the imprinted wave function, while the \textbf{d-l)} show the free time evolution and clustering process of the vortices initially created along the tile edges (see \textbf{c)}). The simulation considers long-living polaritons (conservative case), and the full video is available online from section \ref{sec: links original sims}.
    }
\end{figure}

\newpage

\newpage

\section{Simulations for Tile-8 imprinting Scheme and different tile block size}

\begin{figure}[h!]
    \begin{center}
        \includegraphics[width=0.8\linewidth]{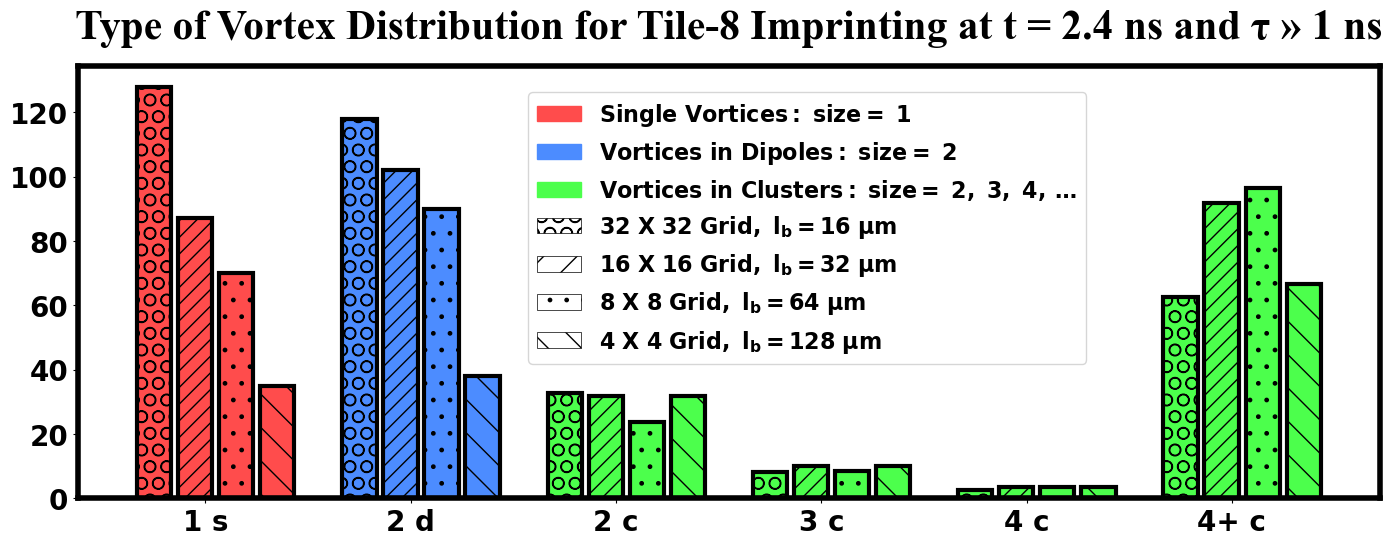}
    \end{center}
    \caption{
    Vortex-statistics histogram comparing the number of solo vortices, vortices in dipoles and vortices in clusters according with cluster size for Tile-8 methods over different tile dimensions. Note that the statistics is computed at a few picoseconds decades after the quantum turbulence analysis considered begins ($t=2.28$~ns). 
    }
\end{figure}

\begin{figure}[h!]
    \begin{center}
        \includegraphics[width=0.8\linewidth]{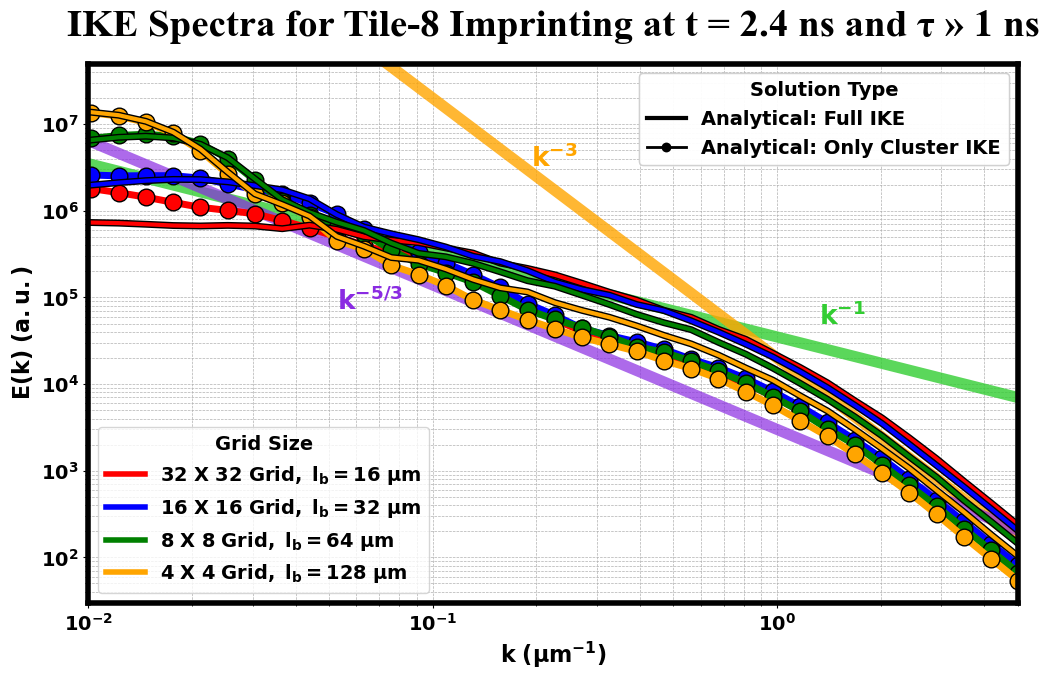}
    \end{center}
    \caption{Incompressible kinetic energy (IKE) spectra comparison for Tile-8 methods over different tile dimensions, a few picoseconds after the quantum turbulence analysis considered begins ($t=2.28$~ns). For each method, the IKE spectrum is analytically computed for both the entire vortex distribution and only-clustered vortex distribution. 
    }
\end{figure}

\begin{figure}[h!]
    \begin{center}
        \includegraphics[width=0.49\linewidth]{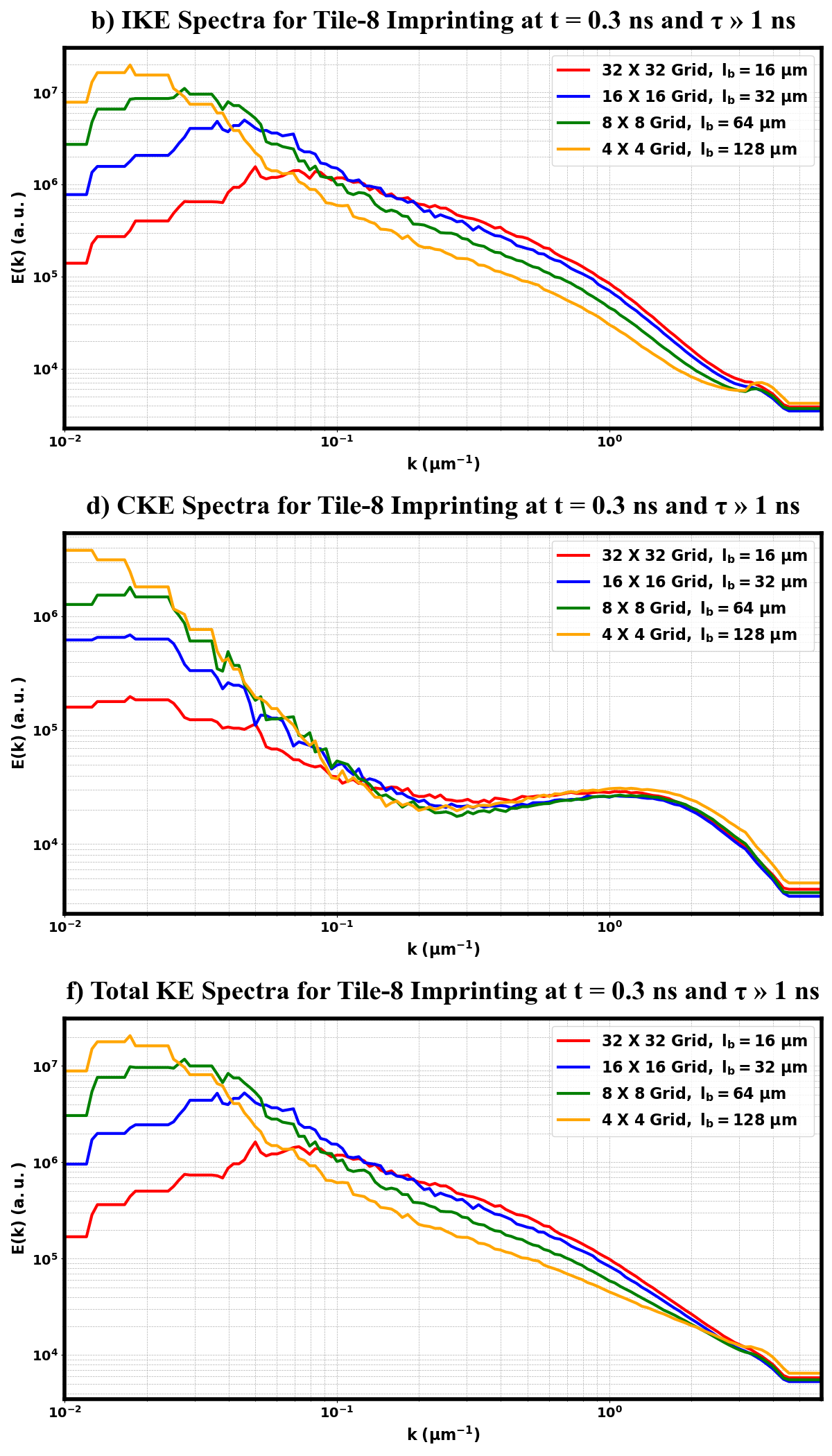}\includegraphics[width=0.49\linewidth]{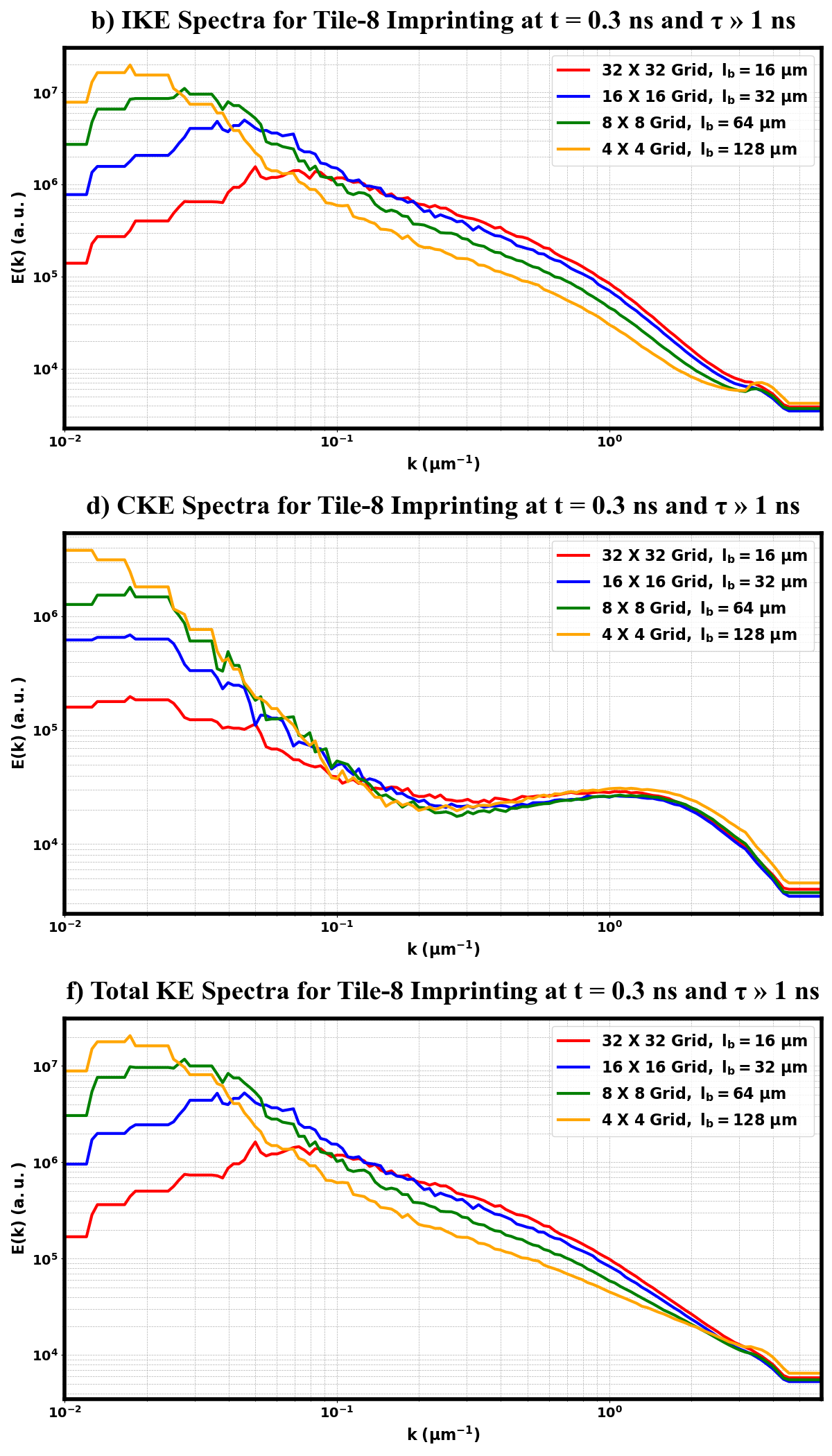}
    \end{center}
    \caption{\label{fig_suppl_stong_depletion} Comparison between incompressible (IKE), compressible (CKE) and total (KE) kinetic energy spectra  for Tile-8 methods over different tile dimensions, at $t=30$~ps and $t=300$~ps, (initial simulation frames after the phase pattern imprinting during $\beta=0.001$ preparatory stage). For each method, the different energy spectra are numerically computed by means of Fourier domain.}
\end{figure}

\newpage

\section{Wave function at strong depletion}

\begin{figure}[h!]
    \begin{center}
        \includegraphics[width=0.8\linewidth]{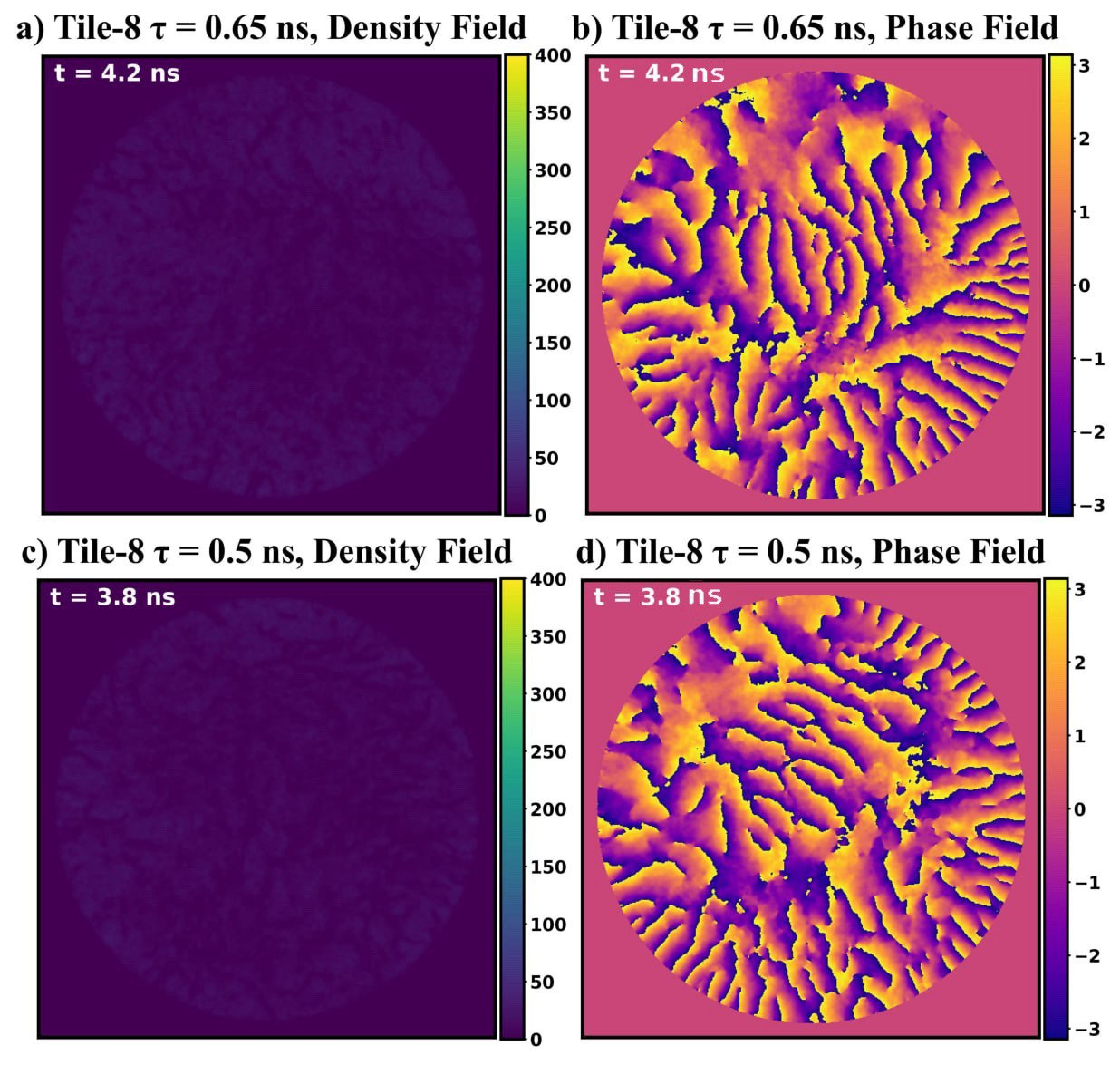}
    \end{center}
    \caption{\label{fig_suppl_stong_depletion} Simulation snapshots for Tile-8 imprinting scheme ($8\times8$ tile and $l_b = 64~\mu$m) for polaritons undergoing a  strong radiative decay during the analysis stage, 2.28~ns $\leq$ 4.56~ns. For lifetime values $\tau=0.65$~ns and $\tau=0.5$~ns, corresponding to density rates $\langle|\psi|^2\rangle(t=4.56~ns)/\langle|\psi|^2\rangle(t=2.28~ns) = 0.03,\ 0.01$ respectively, the condensate depletes so quickly to significantly compromise quantum turbulence and the ambiguous vortex clustering process much before the simulations ends. This is clearly visible in panels \textbf{a,c) }. 
    }
\end{figure}

\end{document}